\documentclass[]{pasj02}
\Received{}%{yyyy/mm/dd}
\Accepted{}%{yyyy/mm/dd}
\usepackage{amsmath} 
\usepackage{xcolor}
\usepackage[switch,mathlines]{lineno}
\begin{document} 

\title{

Inner structure of cold and warm dark matter halos from particle dynamics

%\LETTERLABEL %%% <-- uncomment for LETTER article  
%\REVIEWLABEL %%% <-- uncomment for REVIEW article  
}

%%% begin:list of authors
% Do NOT capitalize all letters in "textsc".
\author{Yohsuke \textsc{Enomoto}\altaffilmark{1}
% \thanks{Present Address is enomoto@tap.scphys.kyoto-u.ac.jp}
}
\altaffiltext{1}{Department of Physics, Kyoto University, Kyoto 606-8502, Japan}
\email{enomoto@tap.scphys.kyoto-u.ac.jp}

\author{Atsushi \textsc{Taruya}\altaffilmark{2,3}}

\author{Satoshi \textsc{Tanaka},\altaffilmark{2}}

\author{Takahiro \textsc{Nishimichi}\altaffilmark{2,3,4}}

\altaffiltext{2}{Centre for Gravitational Physics and Quantum Information, Yukawa Institute for Theoretical Physics, Kyoto University, Kyoto 606-8502, Japan}

\altaffiltext{3}{Kavli Institute for the Physics and Mathematics of the Universe (WPI),
The University of Tokyo Institutes for Advanced Study, The University of Tokyo,\\ 5-1-5 Kashiwanoha, Kashiwa, Chiba 277-8583, Japan}

\altaffiltext{4}{Department of Astrophysics and Atmospheric Sciences, Faculty of Science,
Kyoto Sangyo University, Motoyama, Kamigamo, Kita-ku, Kyoto 603-8555, Japan}

%%% end:list of authors

%% `\KeyWords{}' always has to be placed before ``\maketitle'' 
%%  List of Key Words:  https://academic.oup.com/pasj/pages/Pasj_Keywords 
\KeyWords{Galaxy: halo --- dark matter --- cosmology: theory --- large-scale structure of universe}  

% \thisfancyput(14.6cm,0.0cm){{\large YITP-24-123}}

\maketitle

\begin{abstract}
Using the number of apocenter passages $p$ and the radial action $J_r$ of each particle, we characterize the phase-space structure within the multi-stream regions of cold and warm dark matter halos in cosmological $N$-body simulations. 
Building on previous work by \citet{Enomoto2024}, we analyze the radial density profiles of particles classified by $p$ and $J_r$. 
We find that the profiles consistently follow a double power-law structure, independent of the dark matter model or halo mass. 
The inner profile exhibits a $\rho \propto r^{-1}$ behavior, which is consistent with previous studies. 
Notably, this characteristics persist across both classification schemes. 
In contrast, the outer power-law profiles display distinct behaviors depending on the classification. 
While particles classified by $p$ exhibit a steeper slope, ranging from $-6$ to $-8$, those classified by $J_r$ follow a common slope of approximately $-3.5$.
Overall, the amplitude of the double power-law profiles varies between simulations for different dark matter models, but this variation can be attributed to statistical differences in the concentration of halos across the models.
\end{abstract}
% \pagewiselinenumbers

\section{Introduction}\label{sec:introduction}
In the concordance model of our universe, i.e., Lambda Cold Dark Matter ($\Lambda$CDM) model, the major building blocks in the structure formation of the universe are the dark matter halos, which are self-gravitating objects collapsed from initial density fluctuations.
Cosmological $N$-body simulations have been playing a major role in providing robust predictions of the structure of halos that are the basis for observationally verifying dark matter models.
They have successfully reproduced the observed large-scale structures of the universe such as galaxy distribution at the local universe (e.g., \cite{Springel2006Natur}).

One important prediction of $N$-body simulations is the universal radial density profile of CDM halos irrespective of their masses, which is called the Navarro-Frenk-White (NFW) profile \citep{Navarro1996ApJ, Navarro1997, Navarro2010MNRAS, Wang2020Natur}.
Because of the cold and collisionless nature of CDM, the collapse of initial density fluctuation is described as a folding three-dimensional sheet of dark matter distribution in a six-dimensional phase space, which follows the Vlasov-Poisson equation (e.g., \cite{Colombi2021}).
Hence the internal structure of halos is multi-folded three-dimensional sheets with multi-valued velocity flows at a given position, it is called a multistream region (\cite{More2015ApJ}, see also \cite{Diemer2014ApJ}).
The universality of the NFW profile is an inherent feature of the multistream region of CDM halos.
Furthermore, the multistream region exhibits another universality: the pseudo-phase space density follows a simple power law of radius \citep{Taylor2001ApJ, Ludlow2011MNRAS}.

Despite of many works that have tried to explain the origin of these universalities via different types of arguments, the theoretical and concordant understanding of them from fundamental principles is still missing.
\citet{Syer1998MNRAS} and \citet{Nusser1999MNRAS} discussed that the central cusps of CDM halos are created from the hierarchical clustering of substructures and the index of the slope can be described by the spectral index of the linear power spectrum (as early studies on the central cusps, see also, e.g., \citet{Fukushige1997ApJ, Kravtsov1998ApJ, Moore1999MNRAS, Jing2000ApJ, Subramanian2000ApJ, Ricotti2003MNRAS, Fukushige2004ApJ}).
While \citet{Wang2007} demonstrated that the NFW profile also appears in halos grown without hierarchical clustering, recent studies show that the merger processes are responsible for evolving the central density profile towards universality \citep{Ogiya2016MNRAS, Angulo2017, 2023DelosWhite}.
Another line of research derives the density profile of halos by extending the maximum entropy theory first advocated by \citet{Lynden-Bell1967MNRAS} to understand the apparent insensitivity of the collisionless self-gravitating system upon initial condition (\cite{Hjorth2010ApJ, Pontzen2013}, see also \cite{Williams2010ApJIII, Williams2014ApJIV, Williams2022ApJV}).
Compared with the density profile, the origin of the universality of the slope of the pseudo-phase space density is unclear or even considered to be a fluke (\cite{Arora2020ApJ}, see also \cite{Lapi2011ApJ, Nadler2017MNRAS, Colombi2021}), although the power law behavior is observationally confirmed in galaxies and galaxy clusters (e.g., \cite{Biviano2013A&A, Chae2014ApJ, Biviano2023A&A}).

Motivated by these, we have recently investigated the phase-space structure of multistream region from particle trajectories \citep{Enomoto2023, Enomoto2024} by extending the method performed in \citet{Sugiura2020} (see also \cite{Diemer2022, Diemer2023MNRAS, Diemer2024arXiv} as similar but different methods).
In these works, by tracking particles' trajectory from $z=5$, we count their number of apocenter passages $p$ to characterize the multi-stream flows and study
the radial density profile of the particles classified by $p$.
Interestingly, we found that the radial density profiles of particles classified by the value of $p$ universally exhibit a double power law whose inner and outer indices of the slope are $-1$ and $-8$ regardless of the halo mass, concentration, or recent mass accretion rate.
We also compared our results with self-similar solutions of spherical collapse models \citep{Fillmore1984ApJ, Sikivie1997PhRvD}, but none of the models could explain the slope of $r^{-1}$ even if we include angular momentum to the models.

Following up on these studies, an investigation of the multistream region of halos in alternative dark matter models might provide a promising clue for distinguishing between dark matter models.
Apart from CDM, several dark matter candidates have been explored motivated by observational facts such as the rotation curve of galaxies or the number of satellite galaxies (see \cite{Bullock_Boylan-Kolchin2017, Angulo2022review} as recent reviews).
One feasible candidate is warm dark matter (WDM), which is thermally created in the early universe and possesses a non-negligible
velocity dispersion corresponding to its mass, leading to a free-streaming cutoff in the linear matter power spectrum.
Due to this cutoff, WDM halos tend to have a lower number of subhalos compared with CDM halos, which is favorable as a solution to the small-sale challenges against the $\Lambda$CDM model (e.g., \cite{Bode2001}).
Therefore, many studies on the structure of WDM halos have been carried out \citep{Viel2005, Villaescusa-Navarro2011, Polisensky2011PhRvD, Lovell2012MNRAS, Maccio2012WDMcore, Maccio2013, Polisensky2014MNRAS, Polisensky2015MNRAS, Leo2017JCAP, Stucker2022, Shtanov2024}, and their density profiles are also well-fitted by the NFW profile \citep{Lovell2014}.
Although the allowed mass range is becoming more constrained, the current limit
on WDM mass ($\gtrsim 5$keV, \cite{Dekker2022, Villasenor2023, Irsic2024, Keeley2024}) remains
several orders of magnitude lower than that of typical CDM mass ($\sim 100$GeV), with further constraints still awaited.

The classification of simulation particles by the number of apocenter passages revealed new universal features in phase space that could serve as the basis for verifying DM models, but calculating $p$ for real astronomical objects such as stars is quite challenging.
To count $p$, we need complete information about
their trajectory from the moment they accreted onto the halo.
This requires a reliable estimate of the past history of the halo's gravitational potential, in addition to the current positions of stars in the $6$D phase space, which is very difficult to obtain.
Hence, from an observational perspective, we need an alternative quantity to characterize the phase-space structure of the multi-stream region.

As a promising candidate with both a strong theoretical foundation and observational significance,
the radial action $J_r$ is a favorable quantity to characterize the phase-space structure of halos.
It is defined by \citep{BinneyTremaine2008}
\begin{eqnarray}\label{eq:def_of_jr}
    J_r &\equiv& \frac{1}{2\pi} \oint v_r(r)dr  \nonumber \\
    &=& \frac{1}{\pi} \int_{r_\mathrm{peri}}^{r_\mathrm{apo}} 
    \sqrt{2E-2\Phi(r;t)-j^2/r^2} dr
\end{eqnarray}
where $\Phi$ is the
gravitational potential at a given radius $r$, and $v_r(r)$, $E$, $j$, $r_\mathrm{peri}$ and $r_\mathrm{apo}$ are the radial velocity as a function of $r$, the specific energy, the angular momentum, the pericenter, and the apocenter of the particle, respectively.
Because radial action is an adiabatic invariant and remains conserved if the potential is static or evolving slowly compared to
the orbital period of the particle, the clustering of stars in action space is used to classify stellar streams (e.g., \cite{Naidu2020ApJ, Arora2022ApJ, Malhan2022ApJ, Brooks2024, Malhan2024ApJ}).
Moreover, it plays a key role as a conserved quantity in connecting the initial linear density field to the radial density profile of the objects that collapse from that initial field (e.g., \cite{Fillmore1984ApJ, Nusser2001MNRAS, Dalal2010}).
In addition, \citet{Pontzen2013} found that the conservation of radial action imposes a constraint on maximizing entropy in self-gravitating systems, significantly improving the prediction of the phase-space distribution function, though their model still fails to describe the inner part of halos.
\citet{Burger2021} also analyzed the distribution of radial action in halos and developed a theoretical model that captures the evolution of radial action in a quasi-equilibrium regime, which basically explains the systematic difference in the distribution of radial action between dark matter only and hydrodynamics simulations found in \citet{Callingham2020MNRAS}.

In this paper, aiming to characterize the multi-stream regions of CDM and WDM halos and take a first step toward observational tests of dark matter models in phase space, we compute both the number of apocenter passages $p$ and the radial action $J_r$ for simulation particles.
Then, we classify the particles in halos by the two variables and analyze their radial density profile.
We find that the profiles exhibit a double power law with the inner slope of $-1$ for both the WDM and the CDM cases.
We further show that the inner profile of the two models has different amplitudes, and it is explained by the difference in the distribution of the concentration parameter of halos between the models.

This paper is structured as follows. We introduce the simulation data in section~\ref{subsec:simulations} and our halo catalog in section~\ref{subsec:halo_catalog}.
In section~\ref{subsec:count_p} and \ref{subsec:count_Jr}, we describe the method for counting the number of apocenter passages and estimating the radial action of simulation particles.
The profiles for individual halos are discussed in section~\ref{subsec:individual}, and stacking analyses follow in section~\ref{subsec:stacked}.
We point out the similarity between the classification by $p$ and $J_r$ in section~\ref{subsec:connection_Jr_p}, and 
% comment on the indices of the inner slope of the profiles in section~\ref{subsec:on_the_slope} (\tnrv{\bf <- Is this subsection gone? No other topics in the discussion section?}), then 
conclude our results in section~\ref{sec:conclusion}.

%\newpage
%%%%%%%%%%%%%%%%%%%%%%%%%%%%%%%%%%%%%%%%%%%%%%%%%%%%%%%%%%%%%%%%%%
%%%%%%%%%%%%%%%%%%%%%%%%%%%%%%%%%%%%%%%%%%%%%%%%%%%%%%%%%%%%%%%%%%
\section{Data}
\label{sec:data}
%%%%%%%%%%%%%%%%%%%%%%%%%%%%%%%%%%%%%%%%%%%%%%%%%%%%%%%%%%%%%%%%%%
%%%%%%%%%%%%%%%%%%%%%%%%%%%%%%%%%%%%%%%%%%%%%%%%%%%%%%%%%%%%%%%%%%

%%--%%--%%--%%--%%--%%--%%--%%--%%--%%--%%--%%--%%--%%--%%--%%--%%
%%--%%--%%--%%--%%--%%--%%--%%--%%--%%--%%--%%--%%--%%--%%--%%--%%
\subsection{$N$-body simulations}\label{subsec:simulations}
%%--%%--%%--%%--%%--%%--%%--%%--%%--%%--%%--%%--%%--%%--%%--%%--%%
%%--%%--%%--%%--%%--%%--%%--%%--%%--%%--%%--%%--%%--%%--%%--%%--%%

%%%%%%%%%%%%%%%%%%%%%%%%%%%%%%%%%%%%%%%%%%%%%%%%%%%%%%%%%%%%%%%%
\begin{table*}
  \tbl{Parameters for $N$-body simulations. $m_\mathrm{DM}$ denotes the dark matter mass, $k_\mathrm{peak}$ the wavenumber at which the dimensionless linear power spectrum $k^3P(k)/2\pi^2$ has its maximum value, $L$ the size of simulation box per side, $N$ the number of simulation particles, $m_p$ the mass of simulation particles, $\epsilon$ the softening length, $N_\mathrm{snaps}$ the number of snapshots we stored.}
  {%
  \begin{tabular}{cccccccc}
      \hline
      Name & $m_\mathrm{DM}$ & $k_\mathrm{peak}$ & $L$ & $N$ & $m_p$ & $\epsilon$ & $N_\mathrm{snaps}$ \\ 
       & [keV] &$[h \mathrm{Mpc}^{-1}]$  &$[h^{-1}\mathrm{Mpc}]$ & & $[10^6 h^{-1}M_\odot]$ & $[h^{-1}\mathrm{kpc}]$ & \\
      \hline \hline
      CDMLR & -  & - & 20 & $500^3$ & 5.61 & 1.2 & 1000\\
      WDM1 & 1 & 4.1 & 20 & $500^3$ & 5.61 & 1.2 & 1000\\
      WDM05 & 0.5 & 2.1 & 20 & $500^3$ & 5.61 & 1.2 & 1000\\
      \hline
      CDMHR &  - & - & 20 &$1000^3$ & 0.70 & 0.6 & $1(z=0)$ \\
      \hline
    \end{tabular}
    }
    \label{tab:simulation_setup}
\end{table*}
%%%%%%%%%%%%%%%%%%%%%%%%%%%%%%%%%%%%%%%%%%%%%%%%%%%%%%%%%%%%%%%%

%%%%%%%%%%%%%%%%%%%%%%%%%%%%%%%%%%%%%%%%%%%%%%%%%%%%%%%%%%%%%%%%
\begin{figure}
    \centering
    \includegraphics[width=8cm]{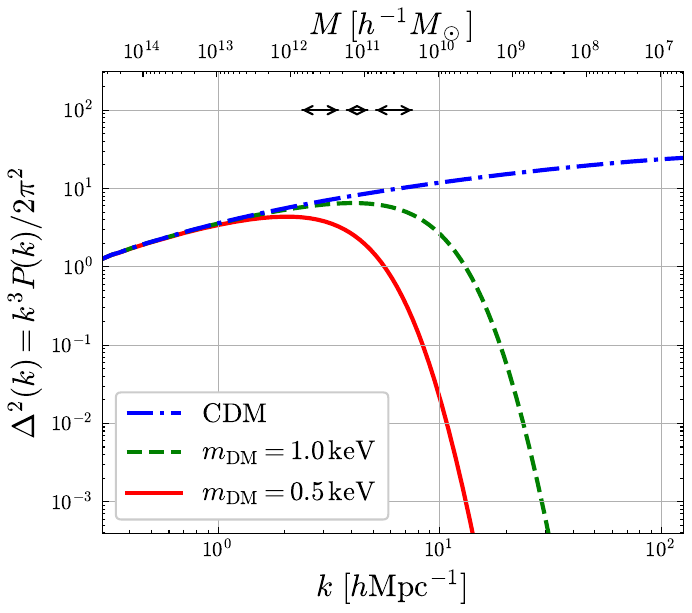}
    \caption{Dimensionless linear power spectrum $\Delta^2(k)$ of our simulations.
    For two WDM cases, we compute $\Delta^2(k)$ by multiplying the transfer function proposed by \citet{Viel2005} (see section~\ref{subsec:simulations}) to that of CDM case computed by CLASS Boltzmann solver \citep{Blas2011}.
    On the upper x-axis, linear mass corresponding to the wavenumber $k$ defined by $M(k) = 4\pi/3 (\pi/k)^3 \bar{\rho}_\mathrm{m}$ \citep{Bullock_Boylan-Kolchin2017} is indicated.
    The horizontal arrows at the panel's top indicate the three mass ranges adopted for stacking analysis.
    Alt Text: Linear power spectra of our dark matter models, focusing especially on the free-streaming scale of the warm dark matter models, where the difference between the models is visible for the range of halo mass we considered.}
    \label{fig:power_spectrum}
\end{figure}
%%%%%%%%%%%%%%%%%%%%%%%%%%%%%%%%%%%%%%%%%%%%%%%%%%%%%%%%%%%%%%%%

In this paper, we will present the results from the three different types of cosmological $N$-body simulations. One is for the cold dark matter (CDM), and the other two simulations are for the warm dark matter (WDM), having the mass of $0.5$ and $1\, \mathrm{keV}$. In all cases, we adopt a periodic comoving box with a size of $20\,h^{-1} \mathrm{Mpc}$ per side and the number of dark matter particles of $N=500^3$, assuming a flat $\Lambda$CDM universe consistent with the cosmological parameters determined by the Planck satellite:  $\Omega_m = 0.3156$, $\Omega_\Lambda=0.6844$, $h=0.6724$, $n_s=0.9645$ and $A_s=2.1307\times10^{-9}$
% and $M_\nu = 0$ eV 
\citep{Planck2015}. 
In addition, we also run another CDM simulation dubbed CDMHR, whose resolution per side is twice higher than those of the three simulations to determine the convergence radii of the density profiles in terms of the softening length. Table~\ref{tab:simulation_setup} summarizes the specification of parameter setup for these simulations. Note that the current observational bounds on the mass of WDM are around $m_{\rm DM}\simeq 5 \mathrm{keV}$ \citep{Dekker2022, Villasenor2023, Irsic2024, Keeley2024}, which is larger than the models we consider. Our primary purpose is to elucidate the dependence of the internal halo structure on the initial conditions as well as to understand better the physical origin of {\it universal} features found by \citet{Enomoto2023}. Thus, even the analysis with disfavored WDM models would still give valuable insights and a clue to discriminate between DM models. 

To run the simulations, the initial conditions are generated from linear power spectra corresponding to different dark matter masses. Simulation particles are displaced from a regular lattice using second-order Lagrangian perturbation theory (2LPT; \cite{Scoccimarro1998, Crocce2006}) to set the initial positions and velocities. We use the CLASS Boltzmann solver \citep{Blas2011} to obtain the linear power spectrum for the CDM case. On the other hand, for the two WDM cases, we adopt the transfer function provided by \citet{Viel2005}, with the functional form originally proposed by \citet{Bode2001}, and calculate their linear power spectra by multiplying it by the linear power spectrum of CDM:
%%%%%%%%%%%%%%%%%%%%%%%%%%%%%%%%%%%%%%%%%%%%%%%%%%%%%%%%%%%%%%%%%%%%%%%
\begin{eqnarray}
    T(k) &\equiv& \sqrt{P_\mathrm{WDM}(k)/P_\mathrm{CDM}(k)} \nonumber \\
    &=& [1+(\alpha k)^{2\nu}]^{-5/\nu}
\end{eqnarray}
%%%%%%%%%%%%%%%%%%%%%%%%%%%%%%%%%%%%%%%%%%%%%%%%%%%%%%%%%%%%%%%%%%%%%%%
 with the index $\nu$ set to $\nu=1.12$. The coefficient $\alpha$ is approximated by
%%%%%%%%%%%%%%%%%%%%%%%%%%%%%%%%%%%%%%%%%%%%%%%%%%%%%%%%%%%%%%%%%%%%%%%
\begin{equation}
    \alpha = 0.049 \left(\frac{m_\mathrm{DM}}{1 \mathrm{keV}} \right)^{-1.11} 
    \left(\frac{\Omega_\mathrm{DM}}{0.25} \right)^{0.11} \left(\frac{h}{0.7} \right)^{1.22}
    h^{-1} \mathrm{Mpc}    . 
\end{equation}
%%%%%%%%%%%%%%%%%%%%%%%%%%%%%%%%%%%%%%%%%%%%%%%%%%%%%%%%%%%%%%%%%%%%%%%
Since the WDM power spectrum generated by the CLASS Boltzmann solver exhibits numerical oscillations at the relevant scales, we use this fitting function instead. To facilitate comparison between models, we adopt the same Fourier phases for the Gaussian random fluctuations that seed the 2LPT. Figure~\ref{fig:power_spectrum} shows the linear power spectra for the models considered in this paper. We then evolve the initial density field using a TreePM code GINKAKU (Nishimichi, Tanaka \& Yoshikawa in prep.).

In WDM simulations, thermal velocity should, in principle, be included for simulation particles to account for the formation of a thermal core at the centers of halos (e.g., \citet{Hogan2000PhRvD}). 
However, for the 0.5 keV case, the typical thermal velocity is approximately $\sim 0.03 \mathrm{km/s}$ at $z=0$ \citep{Bode2001}, which is negligible compared to the typical minimum velocity in halos ($\sim 1 \mathrm{km/s}$). Furthermore, the thermal core size is around $\sim 0.1 \mathrm{kpc}$ in $0.5$ keV case \citep{Maccio2012WDMcore}, smaller than the softening length of our simulations. 
Consequently, we expect that incorporating this effect would not significantly alter our results and therefore do not include thermal velocity in the simulation particles for WDM cases.

Following \citet{Enomoto2024}, we store 1000 snapshots from $z=5$ to $0$ to track the trajectories of particles and accurately count the number of apocenter passages (see section~\ref{subsec:count_p}).
While \citet{Enomoto2024} sampled snapshots uniformly in redshift, we sample them uniformly in physical time to better track trajectories at lower redshifts, when the orbital periods of particles deep inside the virial radius of halos become significantly shorter.

%%--%%--%%--%%--%%--%%--%%--%%--%%--%%--%%--%%--%%--%%--%%--%%--%%
%%--%%--%%--%%--%%--%%--%%--%%--%%--%%--%%--%%--%%--%%--%%--%%--%%
\subsection{Halo catalog}\label{subsec:halo_catalog}
%%--%%--%%--%%--%%--%%--%%--%%--%%--%%--%%--%%--%%--%%--%%--%%--%%
%%--%%--%%--%%--%%--%%--%%--%%--%%--%%--%%--%%--%%--%%--%%--%%--%%

%%%%%%%%%%%%%%%%%%%%%%%%%%%%%%%%%%%%%%%%%%%%%%%%%%%%%%%%%%%%%%%%
\begin{figure}
    \centering
    \includegraphics[width=8cm]{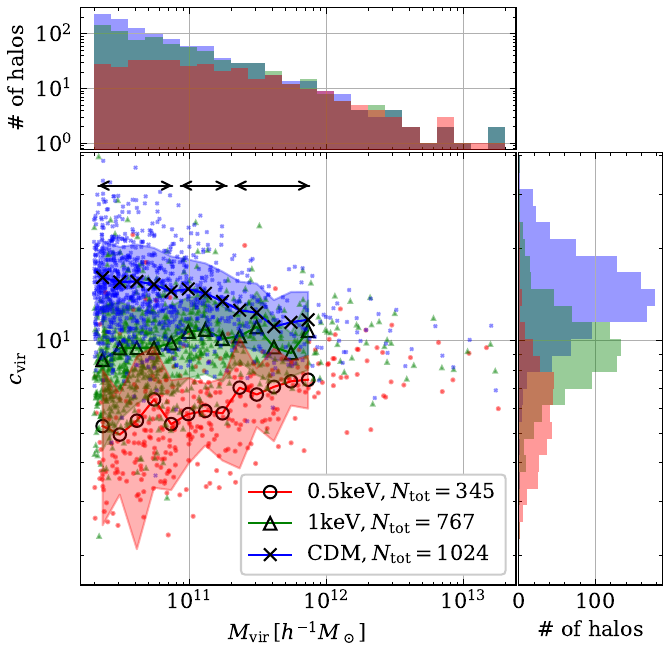}
    \caption{The concentration and mass distribution of our samples (bottom left), and histograms of virial mass (top) and concentration (bottom right).
    Colors correspond to different types of dark matter models indicated in the legend.
    Solid lines and shades depict the mean and standard deviation of concentration at each mass.
    The horizontal arrows indicate the three mass ranges S, M, and L, shown in Table~\ref{tab:halo_catalog}.
    $N_\mathrm{tot}$ is the total number of samples whose mass is larger than $2\times 10^{10}h^{-1}M_\odot$.
    Alt Text: Concentration-mass relation of our halo samples, showing that their characteristics are consistent with the previous studies on WDM models.    
    }
    \label{fig:halo_catalog}
\end{figure}
%%%%%%%%%%%%%%%%%%%%%%%%%%%%%%%%%%%%%%%%%%%%%%%%%%%%%%%%%%%%%%%%

We identify halos at $z=0$ using the 6D phase-space temporal friends-of-friends halo finder ROCKSTAR \citep{Behroozi2013}.
Following \citet{Enomoto2024}, we focus on relaxed halos that meet specific criteria to exclude those undergoing major mergers or classified as subhalos. 
To filter out halos which are likely undergoing major mergers, we apply the following conditions introduced by \citet{Klypin2016}:
%%%%%%%%%%%%%%%%%%%%%%%%%%%%%%%%%%%%%%%%%%%%%%%%%%%%%%%%%%%%%%%%
\begin{equation}\label{eq:crite1}
    \lambda > 0.07, \; X_\mathrm{off}>0.07
\end{equation}
%%%%%%%%%%%%%%%%%%%%%%%%%%%%%%%%%%%%%%%%%%%%%%%%%%%%%%%%%%%%%%%%
where $\lambda$ is the Bullock spin parameter \citep{Bullock2001ApJ} and $X_\mathrm{off}$ is the offset parameter. Both parameters are derived from ROCKSTAR for each halo, and halos meeting the criteria in equation~(\ref{eq:crite1}) are excluded.

Additionally, we exclude subhalos that satisfy the condition:
%%%%%%%%%%%%%%%%%%%%%%%%%%%%%%%%%%%%%%%%%%%%%%%%%%%%%%%%%%%%%%%%
\begin{equation}\label{eq:crite2}
    M_\mathrm{vir,all} > 1.3 M_\mathrm{vir}
\end{equation}
%%%%%%%%%%%%%%%%%%%%%%%%%%%%%%%%%%%%%%%%%%%%%%%%%%%%%%%%%%%%%%%%
where $M_\mathrm{vir,all}$ is the total mass of simulation particles within $R_\mathrm{vir}$ centered at the halo position as calculated by the ROCKSTAR \citep{Enomoto2024}. Here $R_\mathrm{vir}$ and $M_\mathrm{vir}$ refer to the virial radius and mass, respectively, as computed by ROCKSTAR. Since particles inside subhalos are often partially bound to the host halos rather than the subhalos themselves, the condition $M_{\rm vir,all}\gg M_{\rm vir}$ is typically met. \citet{Enomoto2024} shows that combining equation (\ref{eq:crite1}) with equation (\ref{eq:crite2}) serves as an effective method to exclude subhalos.
%Note we also use the virial velocity defined as $V_\mathrm{vir}\equiv \sqrt{GM_\mathrm{vir}/R_\mathrm{vir}}$ below. 

In addition, to accurately track the main progenitor of halos and define continuous trajectories among snapshots for them (see section~\ref{subsec:count_p}), the conditions mentioned above are not still sufficient to remove exceptional halos that are currently undergoing a major merger, having a significant secondary density peak. Hence, 
we impose an additional condition on top of the conditions above, as discussed in \citet{Enomoto2024}:
%%%%%%%%%%%%%%%%%%%%%%%%%%%%%%%%%%%%%%%%%%%%%%%%%%%%%%%%%%%%%%%%
\begin{equation}
    |\boldsymbol{x}_\mathrm{h}-\boldsymbol{x}_\mathrm{h,pro}|/R_\mathrm{vir} > 0.1
\end{equation}
%%%%%%%%%%%%%%%%%%%%%%%%%%%%%%%%%%%%%%%%%%%%%%%%%%%%%%%%%%%%%%%%
where $\boldsymbol{x}_\mathrm{h}$ is the halo center identified with the ROCKSTAR, and $\boldsymbol{x}_\mathrm{h,pro}$ is that of its main progenitor, as will be defined below (see section~\ref{subsec:count_p}).
This last condition allows us to exclude a few irregular halos in each simulation after imposing the criteria given at equations~(\ref{eq:crite1}) and (\ref{eq:crite2}).

For WDM cases, due to the lack of small-scale power in the initial conditions, the relative importance of the discreteness of $N$-body particles becomes significant, and can affect the dynamics of halo formation, resulting in many spurious low-mass halos  (\cite{Wang2007}, \cite{Melott2007}, \cite{Power2016}, see also \cite{Stucker2022} and \cite{Liu2023} for recent progress). To mitigate the impact of such a spurious halo, we here remove the halos smaller than the threshold mass $M_{\rm thre}$ introduced by \citet{Wang2007}:
%%%%%%%%%%%%%%%%%%%%%%%%%%%%%%%%%%%%%%%%%%%%%%%%%%%%%%%%%%%%%%%%
\begin{equation}\label{eq:WDMmthre}
    M_\mathrm{thre} = 10.1\bar{\rho}_\mathrm{m} \,d \,k_\mathrm{peak}^{-2},
\end{equation}
%%%%%%%%%%%%%%%%%%%%%%%%%%%%%%%%%%%%%%%%%%%%%%%%%%%%%%%%%%%%%%%%
where $\bar{\rho}_\mathrm{m}$ is the mean matter density, $d$ is the mean interparticle separation of the simulation, and $k_\mathrm{peak}$ is the wavenumber at which the dimensionless linear power spectrum $k^3P(k)/2\pi^2$ has its maximum value, which is listed in Table~\ref{tab:simulation_setup} for our WDM simulations. From the formula given in equation \eqref{eq:WDMmthre}, the threshold mass is given by $M_\mathrm{thre}=2.1 \times 10^9$ and $8.0 \times 10^9 h^{-1}M_\odot$ for WDM1 and WDM05, respectively, below which we found that the number of halos identified is significantly increased. Note, however, that
the criterion in equation \eqref{eq:WDMmthre} cannot exclude all spurious halos,  
as shown in \citet{Lovell2014} (see their figures 9 and 10). We therefore consider 
a more conservative mass threshold of
% another minimum mass and set it to 
$M_\mathrm{min}=2.0\times 10^{10}h^{-1}M_\odot$.
% as our conservative choice. 
Our final catalog considers halos larger than $M_{\rm min}$.

To illustrate the overall characteristics of the CDM and WDM halos in our catalogs, we plot the concentration-mass relation in figure~\ref{fig:halo_catalog}. As shown in the top histogram,  
the number of WDM halos given as a function of $M_\mathrm{vir}$ is suppressed 
relative to that of CDM halos 
% as decreasing the halo mass, 
toward the low mass end,
and the suppression becomes significant as decreasing the mass of dark matter (e.g., \cite{Bode2001, Bose2016}). 
The same trend is also seen in the concentration-mass relation, that is, the $c_{\rm vir}$ obtained from WDM simulations is systematically suppressed as decreasing $m_{\rm DM}$, and this becomes more prominent at the halo masses below the free-streaming scale for WDM simulations (e.g., see figure 4 of \cite{Ludlow2016}). These are all consistent with previous studies and are solely due to the lack of small-scale power in the linear power spectrum of the WDM models, which delays the formation and accretion/merger history of halos (e.g., \cite{Zhao2009}, \cite{Ludlow2016}, see also \cite{Angulo2022review} as recent review).

%%%%%%%%%%%%%%%%%%%%%%%%%%%%%%%%%%%%%%%%%%%%%%%%%%%%%%%%%%%%%%%%%%
%%%%%%%%%%%%%%%%%%%%%%%%%%%%%%%%%%%%%%%%%%%%%%%%%%%%%%%%%%%%%%%%%%
\section{Method}
%%%%%%%%%%%%%%%%%%%%%%%%%%%%%%%%%%%%%%%%%%%%%%%%%%%%%%%%%%%%%%%%%%
%%%%%%%%%%%%%%%%%%%%%%%%%%%%%%%%%%%%%%%%%%%%%%%%%%%%%%%%%%%%%%%%%%

%%%%%%%%%%%%%%%%%%%%%%%%%%%%%%%%%%%%%%%%%%%%%%%%%%%%%%%%%%%%%%%%%%
\begin{figure*}
    \centering
    \includegraphics[width=16cm]{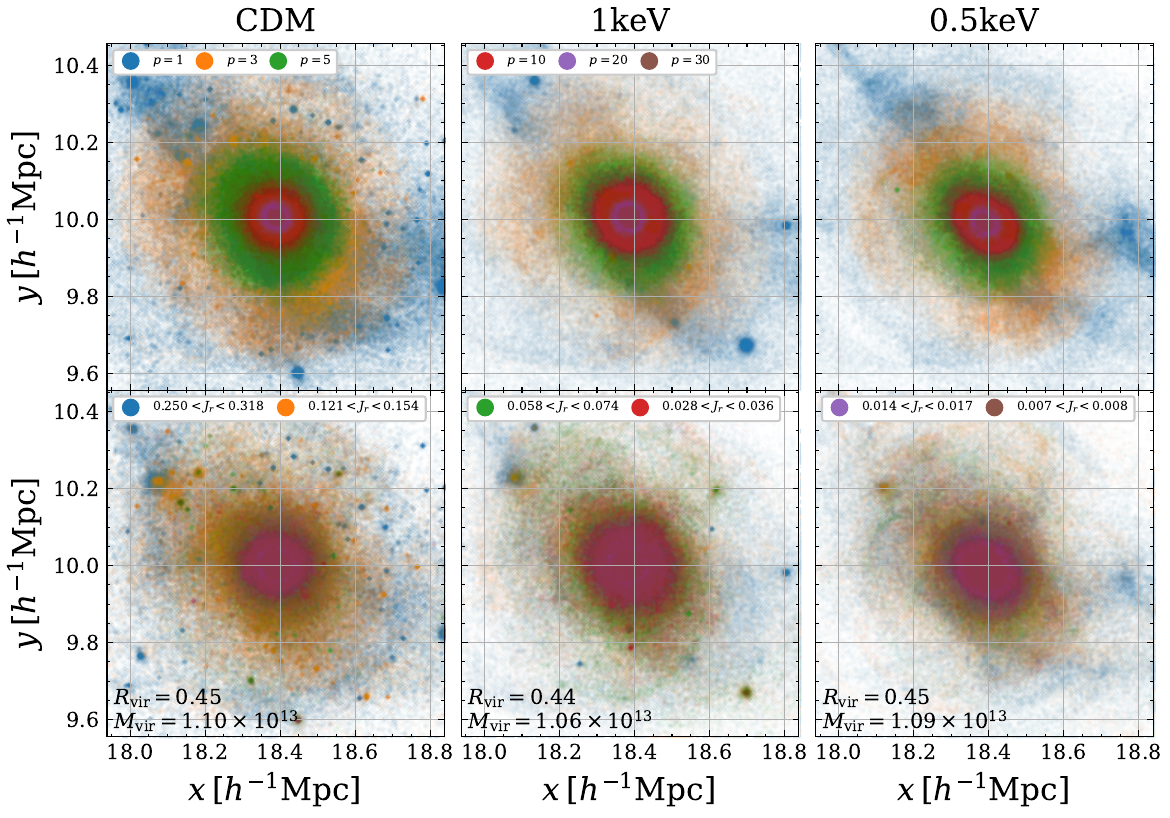}
    \caption{Distribution of particles classified by $p$ and $J_r$ a single halo of the mass $M_{\rm vir}\sim1.1\times10^{13}\,h^{-1}$\,M$_\odot$ with a slice of $0.2\,h^{-1}$Mpc depth for z-axis. 
    
    Particles that are $p=1,3,5,10$ (blue, orange, green, red, accordingly) and those that belong to the four $J_r$ bins indicated in the legend are plotted.
    Virial radius $R_\mathrm{vir}$ and mass $M_\mathrm{vir}$ indicated at the left bottom are in units of $[h^{-1}\mathrm{Mpc}]$ and $[h^{-1}M_\odot]$ respectively.
    $J_r$ shown in the caption is normalized by the unit of $[R_\mathrm{vir}V_\mathrm{vir}]$.
    Alt Text: The snapshots of particles classified by radial action $J_r$ and the number of apocenter passages $p$ in cold and warm dark matter models. Figure shows that the inner part of halos is dominated by the particles with a small value of $J_r$ and large value of $p$.
    }
    \label{fig:pdist_config}
\end{figure*}
%%%%%%%%%%%%%%%%%%%%%%%%%%%%%%%%%%%%%%%%%%%%%%%%%%%%%%%%%%%%%%%%%%

%%%%%%%%%%%%%%%%%%%%%%%%%%%%%%%%%%%%%%%%%%%%%%%%%%%%%%%%%%%%%%%%%%
\begin{figure*}
    \centering
    \includegraphics[width=16cm]{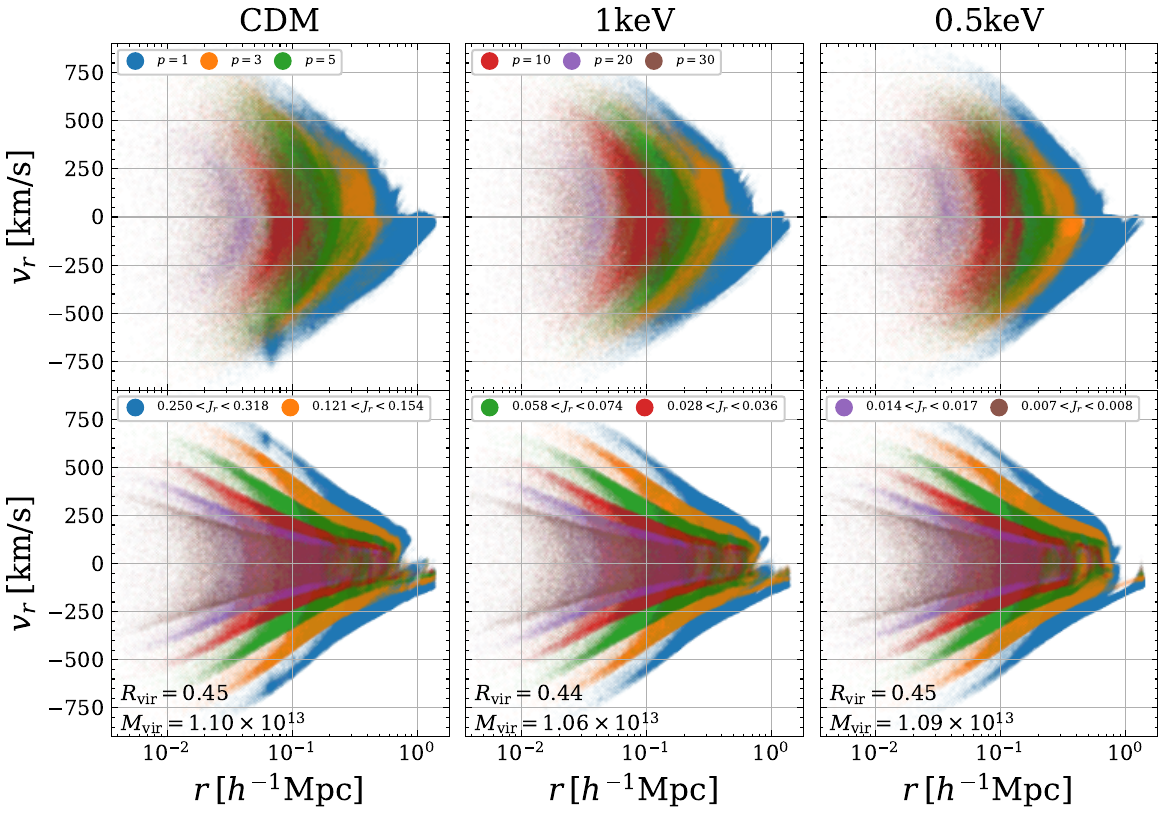}
    \caption{Phase-space distribution of particles classified by $p$ (top row) and $J_r$ (bottom row) in the same halos shown in figure~\ref{fig:pdist_config}. 
    All the particles within the radius $3R_\mathrm{vir}=1.35h^{-1}\mathrm{Mpc}$ are shown, and they are color-coded as same as those in figure~\ref{fig:pdist_config}.
    The leftmost radius shown in each panel is the triple as the softening radius.
    Alt Text: The phase-space snapshots of the representative halos in cold and warm dark matter models. This figure shows that there is no qualitative difference between dark matter models irrespective of the classification of particles by $p$ or $J_r$.
    }
    \label{fig:pdist_phase}
\end{figure*}
%%%%%%%%%%%%%%%%%%%%%%%%%%%%%%%%%%%%%%%%%%%%%%%%%%%%%%%%%%%%%%%%%%

In this paper, we are interested in characterizing the phase-space structures of DM halos based on the information on the motion of DM particles. To be specific, we consider two different methods to quantify the phase-space properties of halos. One is to count the number of apocenter passages of each DM particle, as adopted in our previous works \citep{Enomoto2023,Enomoto2024}. This method was first developed by \citet{Sugiura2020}, extending the algorithm to identify the splashback radius, i.e., the radius of the first apocenter passage, by \citet{Diemer2017ApJS}.
Another is to evaluate the radial action of DM particles \citep{BinneyTremaine2008}. The radial action is frequently used in the subject of galactic dynamics (\cite{Helmi2020review}, \cite{Deason2024review} and \cite{Bonaca2024review} for recent reviews), but it has also been applied to the DM halos in the cosmological context (e.g., \cite{Pontzen2013,Burger2021}). In this section, we describe these two methods in order. 

%%--%%--%%--%%--%%--%%--%%--%%--%%--%%--%%--%%--%%--%%--%%--%%--%%
%%--%%--%%--%%--%%--%%--%%--%%--%%--%%--%%--%%--%%--%%--%%--%%--%%
\subsection{Counting the number of apocenter passages $p$}\label{subsec:count_p}
%%--%%--%%--%%--%%--%%--%%--%%--%%--%%--%%--%%--%%--%%--%%--%%--%%
%%--%%--%%--%%--%%--%%--%%--%%--%%--%%--%%--%%--%%--%%--%%--%%--%%

The number of apocenter passages for each DM particle, which we denote by $p$ in what follows, is a useful quantity to reveal the multi-stream structure of DM flows inside halos. In this paper, we follow the same method as in \citet{Enomoto2024}, using the $1000$ snapshots in $0\leq z\leq 5$. 
In short, we track the center of halos, determined by inner particles, from $z=0$ back to an early time. For each DM particle, we then count the number of apocenter passages, $p$, by following the particle trajectory around the halo center. Finally, we store the value of $p$ at $z=0$.  
Below, we summarize the method described in \citet{Enomoto2024} in more detail.

First, from the halos identified with ROCKSTAR at $z=0$, we track the main progenitor of each halo by following the particles within the virial radii back in time, updating the list of member particles using the shrinking sphere method as described in \citet{Enomoto2024} (see their section~3.1) until the number of member particles falls below 500\footnote{In \citet{Enomoto2024}, this number was set to 1000. We reduced it to 500 so that we could track the halos to an earlier time. Comparing the results with these two parameters, we confirmed that this modification does not affect the results below $p=50$ (see figure A2 of \cite{Enomoto2024}). The results remain unchanged even if we reduce the number of particles from $500$ to $250$.} or the method reaches the first snapshot at $z=5$.
Then, defining the bulk velocity and position of the halos, $\boldsymbol{x}_{\rm h}$ and $\boldsymbol{v}_{\rm h}$ as the average of those 500 particles at each snapshot (i.e., the position and velocity of the main progenitor), we track forward in time and monitor the radial velocity of the particles relative to the center within $2.5 R_\mathrm{vir}$ at $z=0$ from the last snapshot at which we could track the main progenitor.
We count the number of apocenter passages for each particle as the sign of their radial velocity changes from positive to negative, and its position relative to the halo center, $\boldsymbol{x}_{\rm h}$, shifts by at least $90^\circ$ from the previous apocenter passage.

%%--%%--%%--%%--%%--%%--%%--%%--%%--%%--%%--%%--%%--%%--%%--%%--%%
%%--%%--%%--%%--%%--%%--%%--%%--%%--%%--%%--%%--%%--%%--%%--%%--%%
\subsection{Calculating the radial action $J_r$}\label{subsec:count_Jr}
%%--%%--%%--%%--%%--%%--%%--%%--%%--%%--%%--%%--%%--%%--%%--%%--%%
%%--%%--%%--%%--%%--%%--%%--%%--%%--%%--%%--%%--%%--%%--%%--%%--%%

Besides the number of apocenter passages, we also use the radial action $J_r$ to quantify the phase-space structures of DM halos. Following the method presented in \citet{Pontzen2013}, we compute equation~(\ref{eq:def_of_jr}) numerically for each particle at $z=0$. 

In doing so, we first evaluate the spherically averaged potential for each halo defined by
%%%%%%%%%%%%%%%%%%%%%%%%%%%%%%%%%%%%%%%%%%%%%%%%%%%%%%%%%%%%%%%%%%
\begin{equation}
    \Phi(r;t) = -\int_{r}^{\infty} \frac{GM(<r';t)}{r'^2} dr'
\end{equation}
%%%%%%%%%%%%%%%%%%%%%%%%%%%%%%%%%%%%%%%%%%%%%%%%%%%%%%%%%%%%%%%%%%
where $M(<r)$ is the interior mass of the halo enclosed by a sphere of the radius $r$ measured from the halo center ${\boldsymbol x}_{\rm h}$. To compute the integral above, the range of the integral is divided into $[r,\,2.5R_{\rm vir}]$ and $[2.5R_{\rm vir},\infty]$.
While the integral of the former region is performed using the $N$-body data,  the latter region is integrated analytically assuming a Keplerian potential produced by a point particle whose mass is $M(<2.5R_\mathrm{vir})$. 
Here, the choice of the boundary radius is arbitrary, but changing it from $2.5R_\mathrm{vir}$ to $2R_\mathrm{vir}$ varies radial actions for typically only $1\%$ of the particles inside $1.5R_\mathrm{vir}$ by more than $5\%$ from the fiducial value.

Provided the spherically averaged potential, we can compute the specific energy per particle, $E$, defined by $E=v^2/2 + \Phi(r)$, where $v$ is the velocity of each particle subtracting the bulk velocity of halo $v_{\rm h}$. Also, the specific angular momentum $j$ is computed from $j=r v_{\rm t}$, with $v_{\rm t}$ being the tangential velocity, given by $v_{\rm t}=|\boldsymbol{v}-\boldsymbol{v}\cdot {\boldsymbol r}/r|$. Then, using these quantities given as a function of $r$, we can numerically evaluate, for each particle, the integral in equation \eqref{eq:def_of_jr}, where the upper and lower boundary, $r_{\rm apo}$ and $r_{\rm peri}$, are determined by the zero-crossing point of the integrand. 
Note this treatment cannot be applied to the unbound particles at $z=0$ having a positive specific energy. We checked that particles within the virial radius are mostly bounded, and unbound particles are less than $1\%$. 

The method described above only uses the particle data given at $z=0$, assuming that the particle trajectory is determined by the potential measured at $z=0$ and is not affected by other perturbers. Alternatively, using the particle trajectory from the $1000$ snapshot data, we obtain the radial velocity $v_{\rm r}$ as a function of radius for each particle, which can be used to directly compute equation \eqref{eq:def_of_jr}. In appendix~\ref{sec:app1}, we compare both methods and estimate the impact of the method we choose on the inner structure of halos, confirming that the results obtained from stacked halo analysis in section \ref{subsec:stacked} remain insensitive to the choice of the methods. 
In appendix~\ref{sec:app2}, we also checked that our results below are not sensitive to the resolution of $N$-body simulation by comparing our results with those from an additional higher resolution simulation.

%%%%%%%%%%%%%%%%%%%%%%%%%%%%%%%%%%%%%%%%%%%%%%%%%%%%%%%%%%%%%%%%%%
%%%%%%%%%%%%%%%%%%%%%%%%%%%%%%%%%%%%%%%%%%%%%%%%%%%%%%%%%%%%%%%%%%
\section{Results}\label{section:results}
%%%%%%%%%%%%%%%%%%%%%%%%%%%%%%%%%%%%%%%%%%%%%%%%%%%%%%%%%%%%%%%%%%
%%%%%%%%%%%%%%%%%%%%%%%%%%%%%%%%%%%%%%%%%%%%%%%%%%%%%%%%%%%%%%%%%%

In this section, based on the methods described in the previous section, we present the results of phase-space structure inside halos. In section~\ref{subsec:individual}, we first consider the representative halos and show the individual properties of the inner structure of halos. Then, in section~\ref{subsec:stacked}, the stacking analysis is performed, and the radial density profiles are measured for both CDM and WDM halos for a quantitative comparison between DM models.

%%--%%--%%--%%--%%--%%--%%--%%--%%--%%--%%--%%--%%--%%--%%--%%--%%
%%--%%--%%--%%--%%--%%--%%--%%--%%--%%--%%--%%--%%--%%--%%--%%--%%
\subsection{Individual halo properties}\label{subsec:individual}
%%--%%--%%--%%--%%--%%--%%--%%--%%--%%--%%--%%--%%--%%--%%--%%--%%
%%--%%--%%--%%--%%--%%--%%--%%--%%--%%--%%--%%--%%--%%--%%--%%--%%

%%--%%--%%--%%--%%--%%--%%--%%--%%--%%--%%--%%--%%--%%--%%--%%--%%
%%--%%--%%--%%--%%--%%--%%--%%--%%--%%--%%--%%--%%--%%--%%--%%--%%
\subsubsection{Spatial and radial phase-space structures}\label{subsubsec:indiv_conphase}
%%--%%--%%--%%--%%--%%--%%--%%--%%--%%--%%--%%--%%--%%--%%--%%--%%
%%--%%--%%--%%--%%--%%--%%--%%--%%--%%--%%--%%--%%--%%--%%--%%--%%

Here we pick up representative halos from our halo samples and visually look at the similarities and differences of their halos, focusing especially on the spatial and radial phase-space structures.

In figure~\ref{fig:pdist_config}, the spatial distribution of DM around a single halo of the mass $M_{\rm vir}\sim1.1\times10^{13}\,h^{-1}$\,M$_\odot$ is plotted. 
Collecting the particles inside the radius of $3R_{\rm vir}$, the projected particle distribution onto the $x-y$ plane is specifically shown with a slice of $0.2\,h^{-1}$Mpc depth, for CDM (left), WDM of $m_{\rm DM}=1$\,keV (middle), and WDM of $m_{\rm DM}=0.5$\,keV (right). 
Correspondingly, in figure~\ref{fig:pdist_phase}, the radial phase-space distribution of DM particles is shown for the three DM models. 
In both figures, colors in the upper and lower panels indicate the different number of apocenter passages $p$ and different values of radial action $J_r$, respectively. 
Note that, we normalize $J_r$ by $V_{\rm vir}R_{\rm vir}$, where the quantitiy $V_{\rm vir}$ is the virial velocity defined by $V_{\rm vir}=\sqrt{GM_{\rm vir}/R_{\rm vir}}$ here and in the following.

As clearly seen from figure~\ref{fig:pdist_config}, the development of substructures, mainly consisting of subhalos, is different between DM models. While many subhalos are found around the outer part of the halo in the CDM model, the spatial DM distribution gets smooth for WDM models, and the number of subhalo components is significantly decreased when the mass of DM is changed from $m_{\rm DM}=1$\,keV to $0.5$\,keV. Also, arising from different mass accretion histories, the overall shapes of the halos are different.  
These are consistent with those found in previous works (e.g., \cite{Bode2001, Allgood2006}). 

On the other hand, focusing on the spatial pattern of particles classified with $p$ and $J_r$, no notable difference is found among the DM models considered. Typically, particle distribution is segregated with $p$, that is, particles with a larger and smaller value of $p$ tend to reside at the inner and outer part of the halo, respectively. We thus see an onion-like structure in the configuration space (figure~\ref{fig:pdist_config}). Since the halo in both the CDM and WDM models exhibits the multi-stream flow, the onion-like structure can be also seen in the radial phase-space distribution in figure~\ref{fig:pdist_phase}. By contrast, for the radial action $J_r$, the projected distribution in figure~\ref{fig:pdist_config} is mixed up with particles having various values of $J_r$, especially around the central part. This is because even the particles having a specific value of $J_r$ can take various values of specific angular momentum $j$ (see equation~\eqref{eq:def_of_jr}), which allows the motion of DM particles spatially extended, as shown in figure \ref{fig:pdist_phase}.

Since the radial action is considered as an adiabatic invariant, the structure of particle distribution classified with $J_r$ tends to be preserved in a quasi-stationary state. In reality, the quantity $J_r$ gradually decreases with time due to the continuous halo mass growth (see e.g., \cite{Burger2021}), but the particles with a smaller value of $J_r$ would be sufficiently relaxed, erasing the memory of initial conditions. In this respect, one expects that common features can generically appear in both CDM and WDM halos. This would at least hold for the particles with a small value of $J_r$. Since such particles roughly correspond to those with a large value of $p$ near the halo center, a similar common feature can be also expected in the particle distribution classified with $p$, as we have found in the CDM model \citep{Enomoto2023,Enomoto2024}. We will discuss this point in more detail in the following sections.

\subsubsection{Density profiles}\label{subsubsec:indiv_dens}
%%--%%--%%--%%--%%--%%--%%--%%--%%--%%--%%--%%--%%--%%--%%--%%--%%
%%--%%--%%--%%--%%--%%--%%--%%--%%--%%--%%--%%--%%--%%--%%--%%--%%

%%%%%%%%%%%%%%%%%%%%%%%%%%%%%%%%%%%%%%%%%%%%%%%%%%%%%%%%%%%%%%%%%%
\begin{figure*}
    \centering
    \includegraphics[width=16cm]{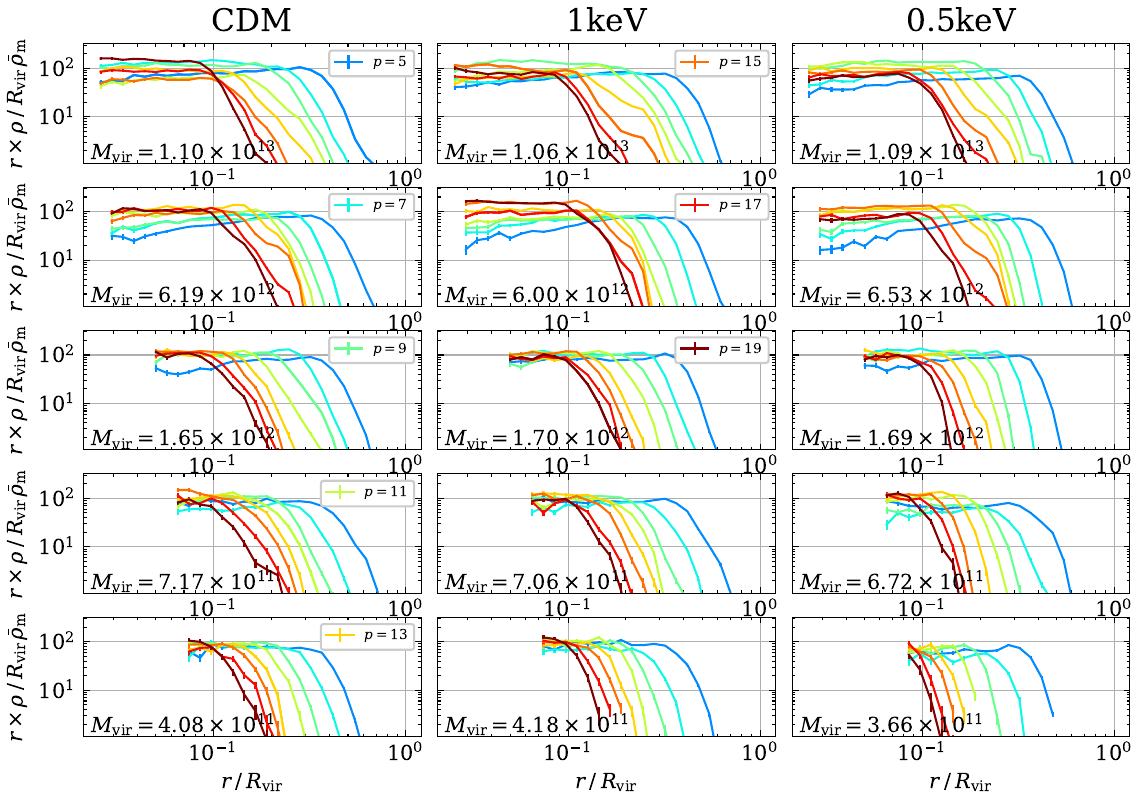}
    \caption{The density profiles for particles classified by the number of apocenter passages in different halos ($p=5,7,9,11,13,15,17,19$, color-coded).
    The error bars are assumed by Poisson scatter, and we only show the profiles in the radii larger than the triple as the softening radius.
    To clearly demonstrate $\rho \propto r^{-1}$, we show $r\times \rho$.
    Most profiles are proportional to $r^{-1}$ for their inner part, except for profiles for $p\lesssim9$ in cluster-sized halos which are supposed to experience a relatively rapid accretion phase currently.
    The three halos in the top row are the same as those shown in figure~\ref{fig:pdist_config}.
    Alt Text: The density profiles of particles classified by the number of apocenter passages, indicating that each profile exhibits double-power law feature as same as CDM results presented in \citet{Enomoto2024}
    }
    \label{fig:pdens_indiv}
\end{figure*}
%%%%%%%%%%%%%%%%%%%%%%%%%%%%%%%%%%%%%%%%%%%%%%%%%%%%%%%%%%%%%%%%%%

%%%%%%%%%%%%%%%%%%%%%%%%%%%%%%%%%%%%%%%%%%%%%%%%%%%%%%%%%%%%%%%%%%
\begin{figure*}
    \centering
    \includegraphics[width=16cm]{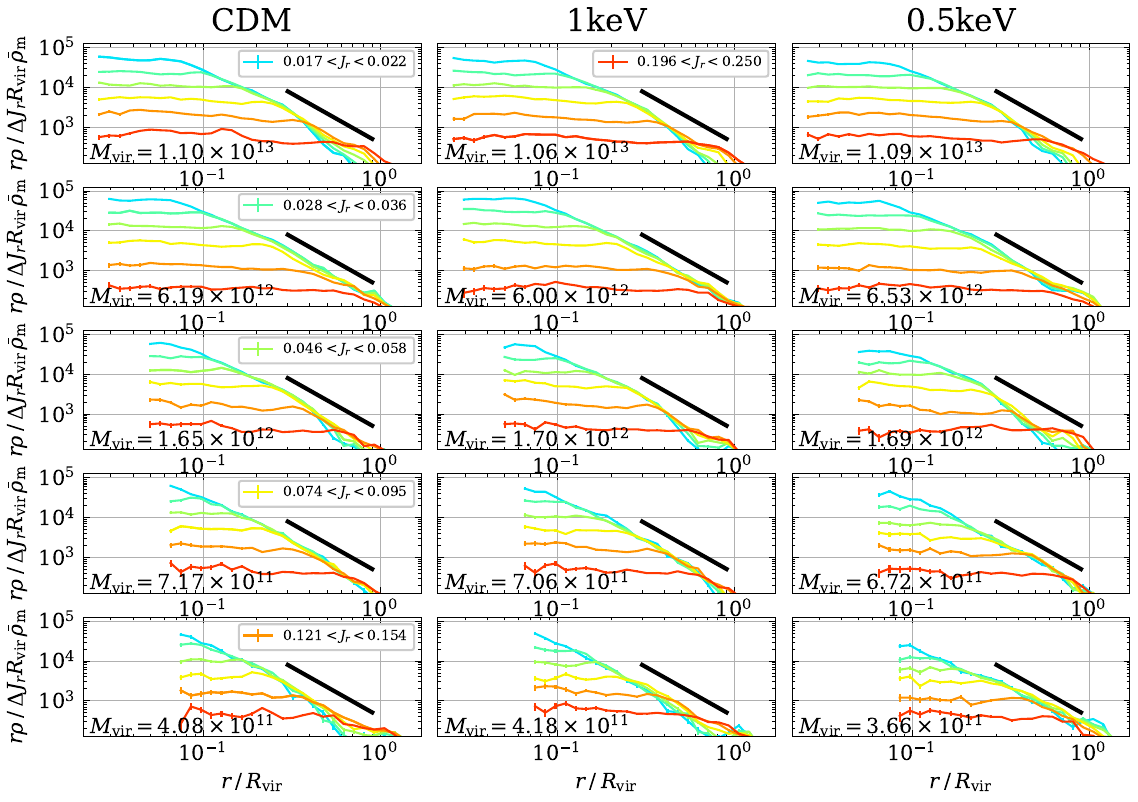}
    \caption{The density profiles for particles classified by radial action $J_r[R_\mathrm{vir}V_\mathrm{vir}]$ in different halos (color-coded).
    Same as figure~\ref{fig:pdens_indiv}, the error bars are assumed by Poisson scatter, and we only show the profiles in the radii larger than the triple as the softening radius.
    We classify particles into log-distributed bins of $J_r$, and divide the profiles by the width of each bin, $\Delta J_r$ as the y-axis indicates.
    The halos shown are the same as those in figure~\ref{fig:pdens_indiv}, and we show $r\times \rho$ again.
    Black solid lines are proportional to $r^{-2.5}$ in this plot.
    Alt Text: The density profiles of particles classified by radial action, indicating that each profile exhibits double-power law feature with indices of the inner slope $-1$ and the outer slope $-3.5$.
    }
    \label{fig:Jrdens_indiv}
\end{figure*}
%%%%%%%%%%%%%%%%%%%%%%%%%%%%%%%%%%%%%%%%%%%%%%%%%%%%%%%%%%%%%%%%%%

To see more quantitatively the inner structure of the halo, we now compute the radial density profiles for particles classified with the number of apocenter passages $p$ and radial action $J_r$. 
In doing so, we select $5$ individual halos identical among the three DM models, including those shown in figures~\ref{fig:pdist_config} and \ref{fig:pdist_phase}. 
We plot in figures~\ref{fig:pdens_indiv} and \ref{fig:Jrdens_indiv} the density profiles as a function of the radius normalized by $R_{\rm vir}$, with the vertical axis multiplied by $r$ and divided by $R_{\rm vir}\overline{\rho}_{\rm m}$. 
Here, in figure~\ref{fig:pdens_indiv}, the results are plotted for particles with an odd number of $p$ out of $5\leq p\leq19$. For the results classified with $J_r$ in figure~\ref{fig:Jrdens_indiv}, we first divide particles into $20$ logarithmic $J_r$ bins at $0.01/\pi \leq J_r/(R_{\rm vir}V_{\rm vir})\leq1/\pi $, and compute the radial density profiles taken from the 6 representative bins out of the range $0.017\leq J_r/(R_{\rm vir}V_{\rm vir})\leq0.25$, where the estimated values of $J_r$ show a good convergence between the two methods described in section~\ref{subsec:count_Jr} (see Appendix \ref{sec:app1}). 
Note that in both figures~\ref{fig:pdist_config} and \ref{fig:pdist_phase}, errorbars in the radial density profiles indicate the Poisson error estimated from the number of particles in each radial bin. 

Overall, the radial profiles shown in both figures~\ref{fig:pdens_indiv} and \ref{fig:Jrdens_indiv} exhibit a double power-law feature, consisting of the inner shallow cusp and the outer steep slope. In particular, we find that the inner profiles commonly have the slope of $-1$, i.e., $\rho\propto r^{-1}$. A closer look at figure~\ref{fig:pdens_indiv} reveals that the inner slope slightly varies with $p$, and tends to have shallower than $-1$ for a small value of $p$. Nevertheless, we see that the slope converges well to $-1$ for $p\geq15$. These are indeed consistent with our previous findings in CDM simulations \citep{Enomoto2023,Enomoto2024}. 
A notable new finding is that these trends persist in the WDM models, even close to the halo mass corresponding to the cutoff scales of the initial power spectrum, e.g., $M_{\rm vir}\sim 10^{12}\,h^{-1}\,M_\odot$ for $m_{\rm DM}=0.5$\,keV (lower right panel of figure~\ref{fig:pdens_indiv}). 

On the other hand, for the profiles of the particles classified with $J_r$ in figure~\ref{fig:Jrdens_indiv}, all of the plotted results show the inner slope of $-1$, and there is no systematic trend with $J_r$. Interestingly, in contrast to those classified with $p$, the outer profiles also show a feature having a common slope close to $-3.5$, as indicated by thick black lines. The trend is particularly prominent for profiles with smaller values of $J_r$, although a small fluctuation is manifest near the virial radius, perhaps due to the non-stationary infalling matter. These results are in fact what is anticipated from the discussions in the previous subsection (see the last paragraph).

\subsection{Stacked halo profiles}\label{subsec:stacked}
%%--%%--%%--%%--%%--%%--%%--%%--%%--%%--%%--%%--%%--%%--%%--%%--%%
%%--%%--%%--%%--%%--%%--%%--%%--%%--%%--%%--%%--%%--%%--%%--%%--%%

%%%%%%%%%%%%%%%%%%%%%%%%%%%%%%%%%%%%%%%%%%%%%%%%%%%%%%%%%%%%%%%%
\begin{table}
  \tbl{The number of halos in each mass range, that are used for stacking analysis.
  The second column from the left shows the mass ranges, and the three columns on the right show the number of halos in the three simulations WDM05, WDM1, and CDMLR.}{%
  \begin{tabular}{ccccc}
      \hline
      Name & $M_\mathrm{vir}[10^{10}h^{-1}M_\odot]$ & $N_\mathrm{0.5keV}$ & $N_\mathrm{1keV}$ & $N_\mathrm{CDM}$ \\ 
      \hline
      S & $[2,8]$  & 153 & 476 & 718 \\
      M & $[8,20]$& 76 & 149 & 165  \\
      L & $[20,80]$& 77 & 104 & 104  \\
      \hline
    \end{tabular}
    }
    \label{tab:halo_catalog}
\end{table}
%%%%%%%%%%%%%%%%%%%%%%%%%%%%%%%%%%%%%%%%%%%%%%%%%%%%%%%%%%%%%%%%

In this subsection, taking advantage of a large number of halo samples, we analyze the density profiles of halos averaged over a certain mass range. For this purpose, we restrict our halo samples to those smaller than $M_{\rm vir}=8\times10^{11}\,h^{-1}\,M_\odot$, and divide them into the three mass ranges, S, M, and L, summarized in Table~\ref{tab:halo_catalog} (see also arrows in figures~\ref{fig:power_spectrum} and \ref{fig:halo_catalog}). Then, we rescale the profile of individual halos by $R_{\rm vir}\overline{\rho}_{\rm m}$ in amplitude and by $R_{\rm vir}$ in radius, and stack the results over the halo samples in each mass range \citet{Enomoto2023,Enomoto2024}. 

Below, to avoid any systematics arising from the softening scales, all of the plotted results are restricted to scales larger than the convergence radius defined as follows. First, we measure the total density profile for stacked halos in the CDM model using the high-resolution simulation (CDMHR). Comparing it with the one obtained from the CDMLR simulations, we determine, for each mass bin, the minimum radius above which both of the simulation results agree well with each other by less than $3\%$. This minimum radius is the convergence radius, and in plotting the density profiles, we apply it to the stacked profiles in both CDM and WDM models. Further, we discard the noisy data and do not show the density profiles if the fractional error, estimated from the individual halo profiles, exceeds $80\%$. 

In section~\ref{subsubsec:stacked_pdens}, we first show the stacked density profiles classified with $p$ and discuss mainly the differences between DM models. We then compute the stacked density profiles with different values of $J_r$ in section~\ref{subsubsec:stacked_Jrdens}, paying attention to the apparent model dependence arising from the halo mass concentration.

%%--%%--%%--%%--%%--%%--%%--%%--%%--%%--%%--%%--%%--%%--%%--%%--%%
%%--%%--%%--%%--%%--%%--%%--%%--%%--%%--%%--%%--%%--%%--%%--%%--%%
\subsubsection{Classification by number of apocenter passages $p$}\label{subsubsec:stacked_pdens}
%%--%%--%%--%%--%%--%%--%%--%%--%%--%%--%%--%%--%%--%%--%%--%%--%%
%%--%%--%%--%%--%%--%%--%%--%%--%%--%%--%%--%%--%%--%%--%%--%%--%%

\begin{figure}
    \centering
    \includegraphics[width=8cm]{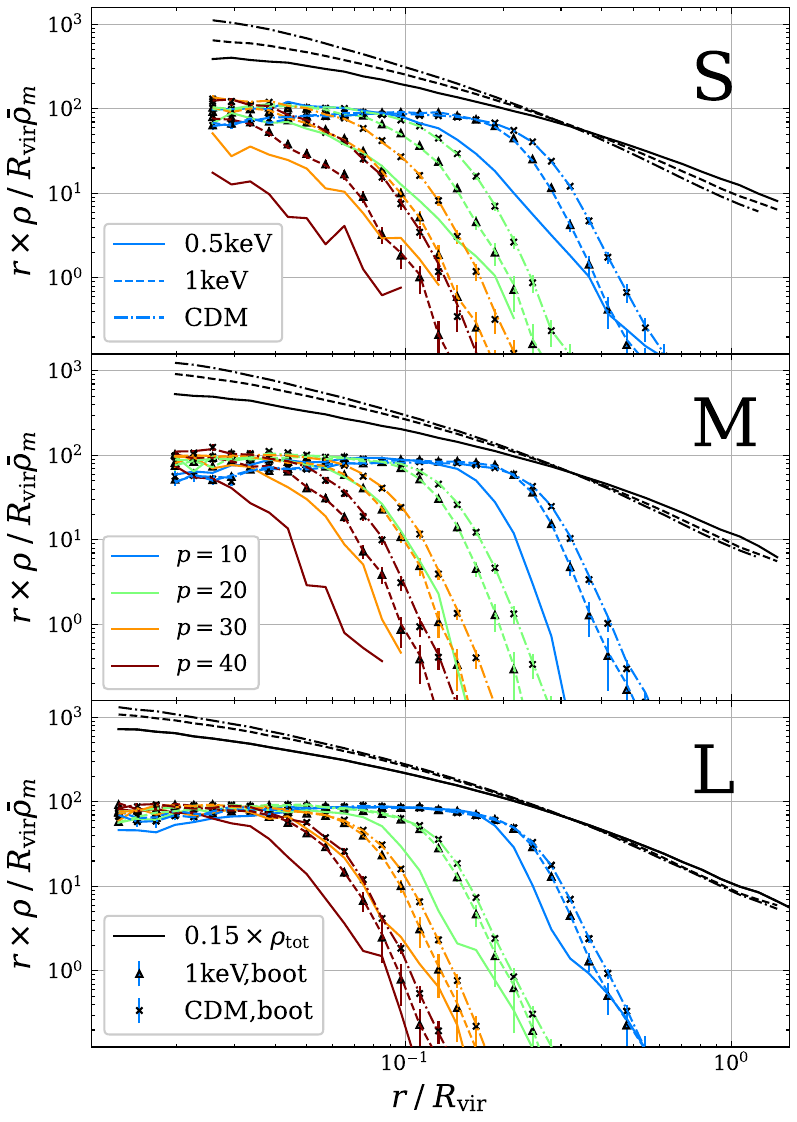}
    \caption{
    The stacked density profiles of particles that are classified by $p=10,15,20,25,30$.
    From top to bottom, we show the results for the mass ranges S, M, and L, and solid, dashed, and dot-dashed lines correspond to $0.5$keV, $1$keV, and CDM simulations respectively.
    Similar to the profiles of individual halos in figure~\ref{fig:pdens_indiv}, stacked profiles are proportional to $r^{-1}$ regardless of the dark matter mass and models.
    As we mentioned in section~\ref{subsec:simulations}, we show the stacked density profiles at the radii where their RMS error are below $80\%$ of their value.
    For clarity, we omit the error bars for the stacked density profiles.
    Alt Text: The stacked density profiles of each mass range, showing each profile exhibit the inner region following $\propto r^{-1}$, and the difference between the dark matter models appears in the outer region of the profiles shown.
    }
    \label{fig:pdens_compar}
\end{figure}

To elucidate the difference between dark matter models, we first consider the DM particles classified with the number of apocenter passages $p$ and show in figure~\ref{fig:pdens_compar} the stacked density profiles for the mass ranges S (upper), M (middle), and L (lower). 
In each panel, we plot the stacked density profiles for particles with $p=10$ (blue), $20$ (green), $30$ (orange) and $40$ (brown). The three different lines represent DM models: 
WDM with $m_{\rm DM}=0.5$\,keV (solid), WDM with $m_{\rm DM}=1$\,keV (dashed), and CDM (dot-dashed). 

Figure~\ref{fig:pdens_compar} clearly shows that profiles classified with $p$ exhibit a double power-law feature with the inner slope of $-1$, irrespective of DM models. Although this is anticipated from the individual halo profile in section~\ref{subsec:individual}, other notable features worth mentioning are summarized as follows:

\begin{description}
    \item[(i)] The inner profiles asymptotically converge to the form $\rho(r;p)\to A(p)\,r^{-1}$ regardless of the DM model, where the amplitude $A(p)$ 
    shows only a weak dependence on halo mass and is approximately $(50-100)\,R_{\rm vir}\overline{\rho}_{\rm m}$. This is particularly the case of a larger value of $p$. 
    \item[(ii)] The characteristic scale where the slope of density profiles changes from $-1$ to a steep value at the outer part systematically moves toward the small radii as we decrease the mass of DM from CDM to WDM with $m_{\rm DM}=0.5$\,keV.  
\end{description}

For the property {\bf(i)}, the weak halo mass dependence of the amplitude $A(p)$ is indeed anticipated from \citet{Enomoto2023,Enomoto2024} for the CDM model. They provided a fitting formula for profiles $\rho(r;p)$ given as the function of $p$ and $M_{\rm vir}$ (see equations~2 and 3 in \cite{Enomoto2023}). 
Although their formula was based on the halo samples with masses in the range $[3.2, 1500]\times 10^{11}\,h^{-1}M_\odot$, 
extrapolating it to the mass range of our halo samples yields the prediction of $A(p)\sim 100-110\,R_{\rm vir}\rho_{\rm m}$ at $p=40$, thus consistent with {\bf(i)}\footnote{The amplitude $A(p)$ here corresponds to the multiple $A^\mathrm{E23}(p)S^\mathrm{E23}(p)$ where $A^\mathrm{E23}(p)$ and $S^\mathrm{E23}(p)$ are the amplitude and the scale radius of the fitting formula $\rho_\mathrm{fit}(r;p)$ obtained in \citet{Enomoto2023} respectively.}. 
%A careful look at the results in the WDM models reveals a subtle difference in $A(p)$, which looks systematically lower than the in the CDM model. 
%\footnote{\cite{Enomoto2024} considered another fitting formula to characterize the double power-law feature for the profiles $\rho(r;p)$, given as a function of the concentration parameter $c_{\rm vir}$ and $p$. They found that the samples classified with $c_{\rm vir}$ exhibit a systematic difference in the amplitude of $A(p)$. Typically, the samples with small $c_{\rm vir}$ have a lower value of $A(p)$. A close look at the differences between DM models, we see that a  
%Since the concentration parameter of the WDM halos is known to be smaller than that of CDM halos, this results are }. 
It is to be noted that the fitting formula actually predicts the amplitude $A(p)$ varying gradually with $p$, and it monotonically decreases. This implies that the convergence of the amplitude $A(p)$ in figure~
\ref{fig:pdens_compar} is superficial, actually changes with $p$, likely for both CDM and WDM halos. Nevetheless, an intriguing point may be the similarity between CDM and WDM halos, with both having the inner slope of $-1$.  

Regarding the feature {\bf(ii)}, a similar trend can be in fact observed if we decrease the number of halos stacked over (see figures 7-10 in \cite{Enomoto2024}). Since the number of halos in WDM models systematically decreases with $m_{\rm DM}$ (see Table~\ref{tab:halo_catalog}), this may be ascribed to the reason for the systematic tendency in the characteristic scale in {\bf(ii)}. 

To clarify this point, for each halo mass bin in the CDM and WDM model with $m_{\rm DM}=1$\,keV, we randomly select halos to match the number of halo samples in the WDM model with $m_{\rm DM}=0.5$\,keV. We compute the stacked density profiles with the reduced number of halo samples. Repeating this procedure $100$ times, we obtain the mean and RMS scatter for the stacked profiles, which are shown in figure~\ref{fig:pdens_compar}, depicted as crosses for CDM and filled triangles for WDM with $m_{\rm DM}=1$\,keV together with errorbars. 

As clearly shown, the resulting profiles nearly coincide with those obtained before reducing the number of samples in every mass bin. Hence the systematic change in the characteristic radius is not artificial, arising from a limited number of halo samples, but inherent, possibly originating from the different formation and mass accretion history among the DM models. Rather, recalling the fact that the concentration parameter of WDM halos becomes smaller than that of the CDM halos (e.g., \cite{Ludlow2016}), the trends seen in figure~\ref{fig:pdens_compar} are consistent with those obtained by \citet{Enomoto2023,Enomoto2024}. In their study, halos in CDM simulations were divided into two halves with the concentration parameter, i.e., one with high- and another with low-concentration halos. They then found that the characteristic scales of the double power-law profiles are systematically smaller for low-concentration halos\footnote{\citet{Enomoto2023,Enomoto2024} also show that the amplitude of the inner density profile for low-concentration halos are systematically lower than that for high-concentration halos (see middle lower panel of fiugre 2 and upper panel of figure 17 of their papers, respectively). Indeed, in figure~\ref{fig:pdens_compar}, we see a systematic difference in amplitude between WDM and CDM halos, although it is subtle and quantitatively small, as  described in the property {\bf(i)}.}, especially for particles with a large value of $p$. This indicates 
a difficulty in discriminating between DM models from the individual halo structure. Rather, the statistical properties of the halo structure like those shown in figure~\ref{fig:pdens_compar} would be important, helping to gain further insight together with the halo concentration properties.

%%%%%%%%%%%%%%%%%%%%%%%%%%%%%%%%%%%%%%%%%%%%%%%%%%%%%%%%%%%%%%%%%%
%%%%%%%%%%%%%%%%%%%%%%%%%%%%%%%%%%%%%%%%%%%%%%%%%%%%%%%%%%%%%%%%%%

\if0 
%%%%%%%%%%%%%%%%%%%%%%%%%%%%%%%%%%%%%%%%%%%%%%%%%%%%%%%%%%%%%%%%%%
\begin{figure}
    \centering
    \includegraphics[width=8cm]{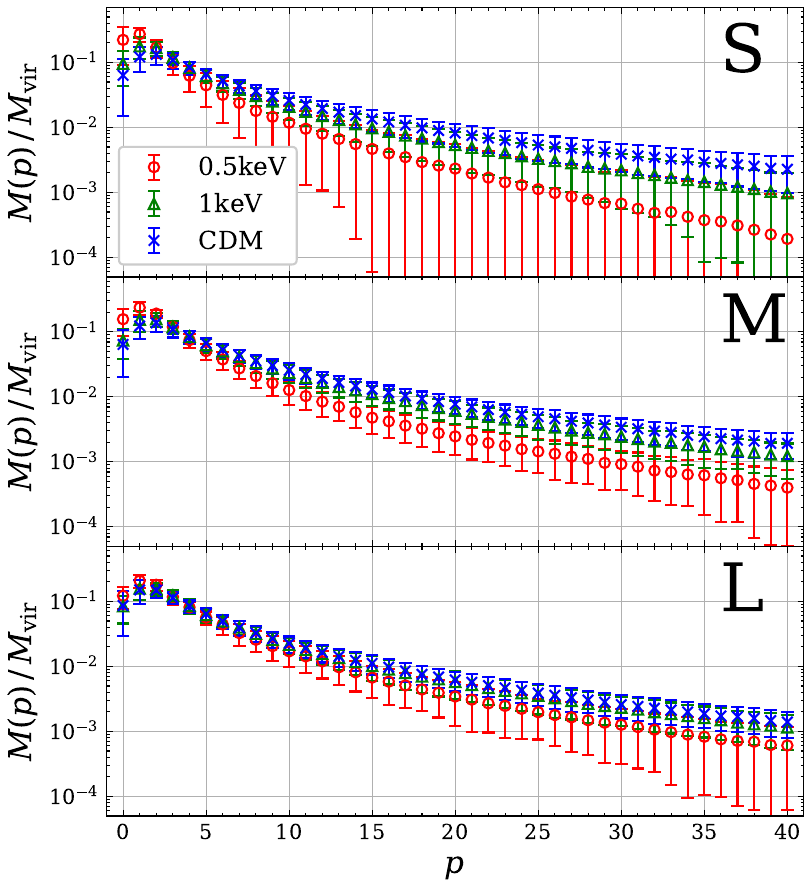}
    \caption{The stacked mass distribution of particles as a function of the number of apocenter passages.
    From the top to the bottom, we show the results for mass ranges S, M, and L.}
    \label{fig:pdistri_all}
\end{figure}
%%%%%%%%%%%%%%%%%%%%%%%%%%%%%%%%%%%%%%%%%%%%%%%%%%%%%%%%%%%%%%%%%%
\fi

%%--%%--%%--%%--%%--%%--%%--%%--%%--%%--%%--%%--%%--%%--%%--%%--%%
%%--%%--%%--%%--%%--%%--%%--%%--%%--%%--%%--%%--%%--%%--%%--%%--%%
\subsubsection{Classification by radial action $J_r$}\label{subsubsec:stacked_Jrdens}
%%--%%--%%--%%--%%--%%--%%--%%--%%--%%--%%--%%--%%--%%--%%--%%--%%
%%--%%--%%--%%--%%--%%--%%--%%--%%--%%--%%--%%--%%--%%--%%--%%--%%

\begin{figure}
    \centering
    \includegraphics[width=8cm]{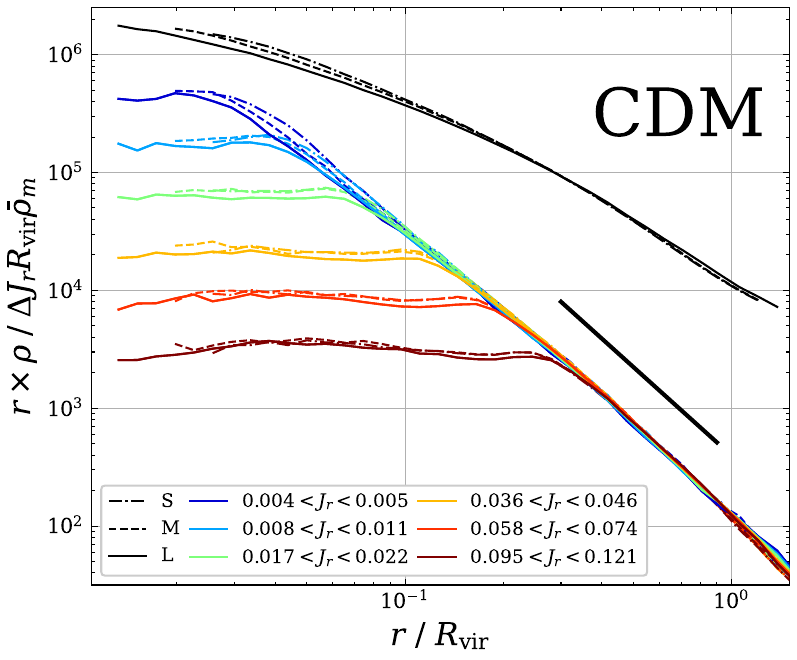}
    \caption{The stacked density profiles of particles in CDM halos that are classified by the radial action $J_r$.
    The top panel shows those for the mass ranges S, M, and L (dot-dashed, dashed, and solid respectively).
    Black solid lines are proportional to $r^{-2.5}$ in this plot.
    Alt Text: The stacked density profiles of particles classified by radial action in the CDM case, showing a clear trend of double-power law feature.
    }
    \label{fig:Jrdens_stacked_CDM}
\end{figure}

Next turn to look at the stacked radial profiles for particles classified with $J_r$. Since this is the first time to characterize them with the radial action quantitatively, we first show the results in the CDM model in figure~\ref{fig:Jrdens_stacked_CDM}. 

The upper panel shows the halo mass dependence of the stacked density profiles classified with $J_r$. In each mass bin, we select 6 representative $J_r$ bins out of the samples divided by the 20 logarithmic $J_r$ bins (see section \ref{subsubsec:indiv_dens}), and plot the results. As it is expected from the individual profiles, the double power-law feature is clearly seen, and for the plotted range of $J_r$, the profiles commonly have the inner and outer slopes of $-1$ and $-3.5$, respectively, the latter of which is indicated by the black thick lines. The characteristic scale where the slope changes from $-1$ to $-3.5$ systematically varies with $J_r$, but both the amplitude and characteristic scale of the profiles do not sensitively depend on the halo mass. Rather, we find that the mass dependence is very weak. A closer look at inner part at $r/R_{\rm vir}\lesssim0.1$ reveals that the dependence tends to become manifest as decreasing $J_r$ and smaller mass halos seem to have a larger amplitude and characteristic scale, although the differences are still small. 

\begin{figure*}
    \centering
    \includegraphics[width=16cm]{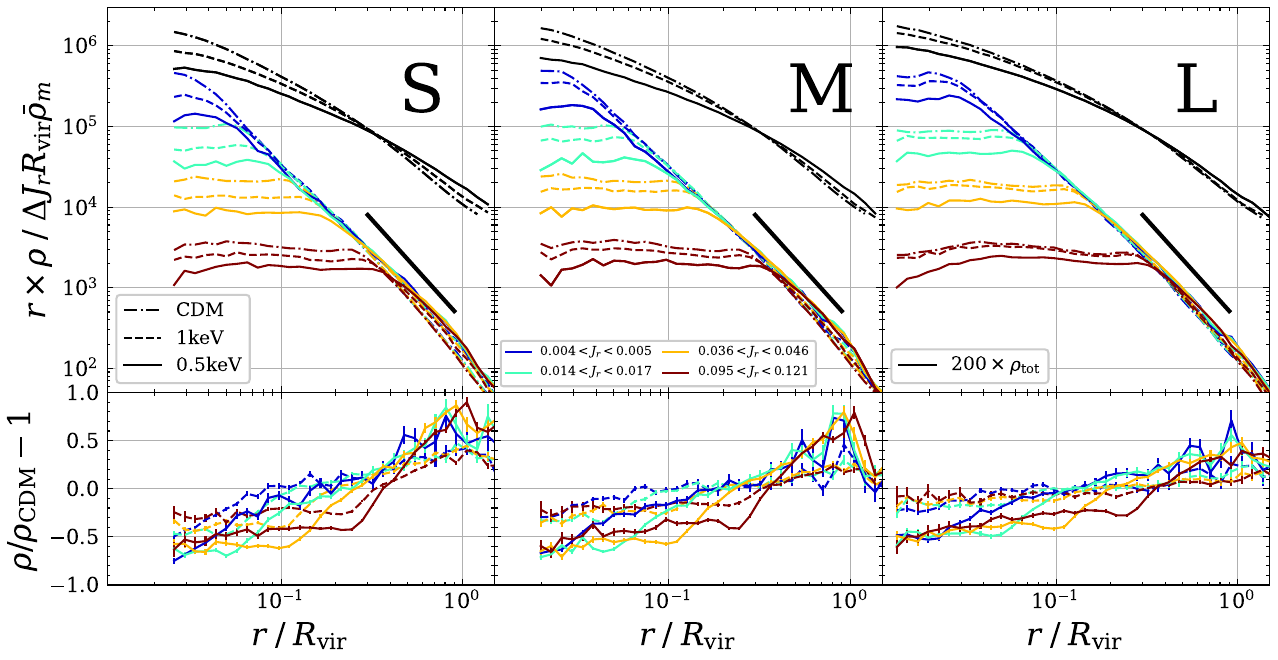}
    \caption{The stacked density profiles of particles in CDM (dot-dashed) and WDM halos (dashed and solid lines for $1$keV and $0.5$keV respectively) that are classified by the radial action $J_r$.
    From left to right, we show the results for the mass ranges S, M, and L.
    In the lower panels, the fractional differences of profiles with respect to the CDM case are shown.
    Alt Text: The stacked density profiles of particles classified by radial action in the CDM and WDM cases. This figure shows a clear difference between the dark matter models especially in the amplitude of the inner profiles, which can be ascribed to the halo concentration between the models.
    }
    \label{fig:Jrdens_stacked_compar}
\end{figure*}

We then look at the WDM models, and plot in  figure~\ref{fig:Jrdens_stacked_compar} their stacked density profiles, together with those in the CDM model. While the upper panels plot the density profiles selected from the representative 4 $J_r$ bins, the bottom panels show the fractional difference of the profile between WDM and CDM models, $\rho_{\rm WDM}/\rho_{\rm CDM}-1$. Overall, the double power-law feature remains universal among the models we consider, with the inner and outer slope of $-1$ and $-3.5$, respectively. In addition, we see the differences between the models in the amplitude and charateristic scales of the density profiles. The WDM halos systematically have lower density at inner radii and higher density at outer radii, compared to those of the CDM halos, and these trends become prominent if the mass of DM is small. Accordingly the characteristic scale of the double power-law feature exhibits the model dependence, and it increases as decreasing the mass of DM. Looking at the halo mass dependence, the differences between the models seems to be significant for smaller halos. 

While these model differences are remarkable and could be a clue to discriminate between DM models from the these profiles, we have seen in section \ref{subsubsec:stacked_pdens} that the differences in the profiles can be originated from the halo concentration properties. Thus, the differences found in figure~\ref{fig:Jrdens_stacked_compar} may be explained by the halo concentration. 

To see this quantitatively, we focus on the halo mass bin S and select halos whose concentration parameters fall within a common range across the models. Specifically, we set the parameter range for $c_{\rm vir}$ to $[4.5,\,6.5]$ for WDM models with mass $m_{\rm DM}=0.5$ and $1.0$\,keV, and $[9.5,\,11.5]$ for the WDM model with mass $m_{\rm DM}=1$\,keV and CDM model.  The number of halos falling into these ranges is around $50-60$ for the former and $80-90$ for the latter. We then compute the stacked density profiles for each sample as similarly shown in figure~\ref{fig:Jrdens_stacked_compar}. 
The results are plotted in figure~\ref{fig:Jrdens_stacked_masscvir}, where we select again the 4 representative $J_r$ bins. We find that the DM model differences mostly disappear (upper panels), and the profiles are very close to each other for each $J_r$ bin. The lower panels plot the fractional differences between the profiles by choosing one of the DM models shown in the upper panels as a reference, i.e., $\rho_{\rm 0.5\,keV}/\rho_{\rm 1\,keV}-1$ for the lower-left panel, and $\rho_{\rm 1\,keV}/\rho_{\rm CDM}-1$ for the lower-right panel. In contrast to those shown in figure~\ref{fig:Jrdens_stacked_compar}, no systematic trend and the model difference is found, and the fractional difference typically lies around $-0.3\sim0.3$, although the variation gets larger as increasing or decreasing the radius, due to a limited number of DM particles. Thus, we conclude that even with the WDM models of the mass $0.5$ and $1$\,keV that are observationally disfavored, the phase-space structure of halos characterized by the radial action has a common feature described by the double power-law profile as seen in the CDM model, with the concentration parameter $c_{\rm vir}$ being the key quantity controlling the amplitude and characteristic scale of the profiles.

%On the other hand, the lower panel of figure~\ref{fig:Jrdens_stacked_CDM} shows the dependence of the concentration parameter. Focusing on the halos of the smallest mass bin S, we divided the samples into two halves, i.e., low- and high-concentration halos, separated by the median value of $c_{\rm vir}=14.9$, and the stacked profiles are plotted. While the double power-law feature is found to be still manifest, we see now a clear systematic difference between the samples, as similarly shown in the cases classified with the number of apocenter passages $p$ \cite{Enomoto2023,Enomoto2024}. That is, the inner amplitude and characteristic scale of the profiles for high-concentration halos are respectively higher and smaller than those for the low-concentration halos, irrespective of $J_r$.   

%%%%%%%%%%%%%%%%%%%%%%%%%%%%%%%%%%%%%%%%%%%%%%%%%%%%%%%%%%%%%%%%%%
%%%%%%%%%%%%%%%%%%%%%%%%%%%%%%%%%%%%%%%%%%%%%%%%%%%%%%%%%%%%%%%%%%

%%%%%%%%%%%%%%%%%%%%%%%%%%%%%%%%%%%%%%%%%%%%%%%%%%%%%%%%%%%%%%%%%%
%%%%%%%%%%%%%%%%%%%%%%%%%%%%%%%%%%%%%%%%%%%%%%%%%%%%%%%%%%%%%%%%%%

\if0
%%%%%%%%%%%%%%%%%%%%%%%%%%%%%%%%%%%%%%%%%%%%%%%%%%%%%%%%%%%%%%%%%%
\begin{figure}
    \centering
    \includegraphics[width=8cm]{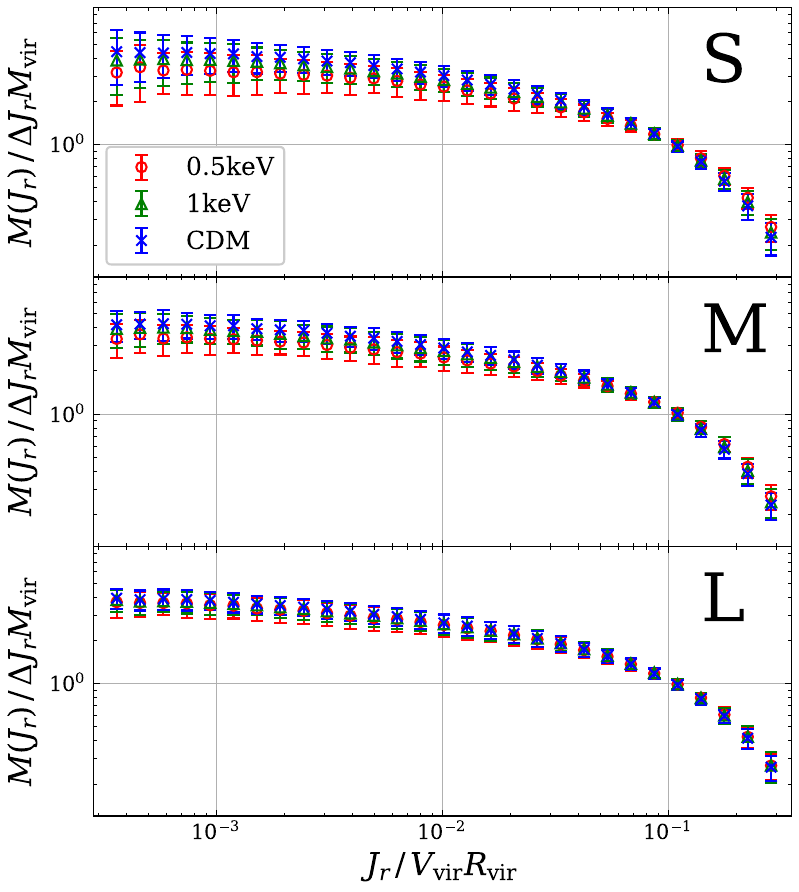}
    \caption{
    The stacked mass distribution of particles as a function of the radial action $J_r$.
    From the top to the bottom, we show the results for mass ranges S, M, and L.}
    \label{fig:Jrdistri_stacked_compar}
\end{figure}
%%%%%%%%%%%%%%%%%%%%%%%%%%%%%%%%%%%%%%%%%%%%%%%%%%%%%%%%%%%%%%%%%%
\fi

%%%%%%%%%%%%%%%%%%%%%%%%%%%%%%%%%%%%%%%%%%%%%%%%%%%%%%%%%%%%%%%%%%
\begin{figure*}
    \begin{tabular}{cc}
        \begin{minipage}{\columnwidth}
        \centering            
        \includegraphics[scale=0.6]{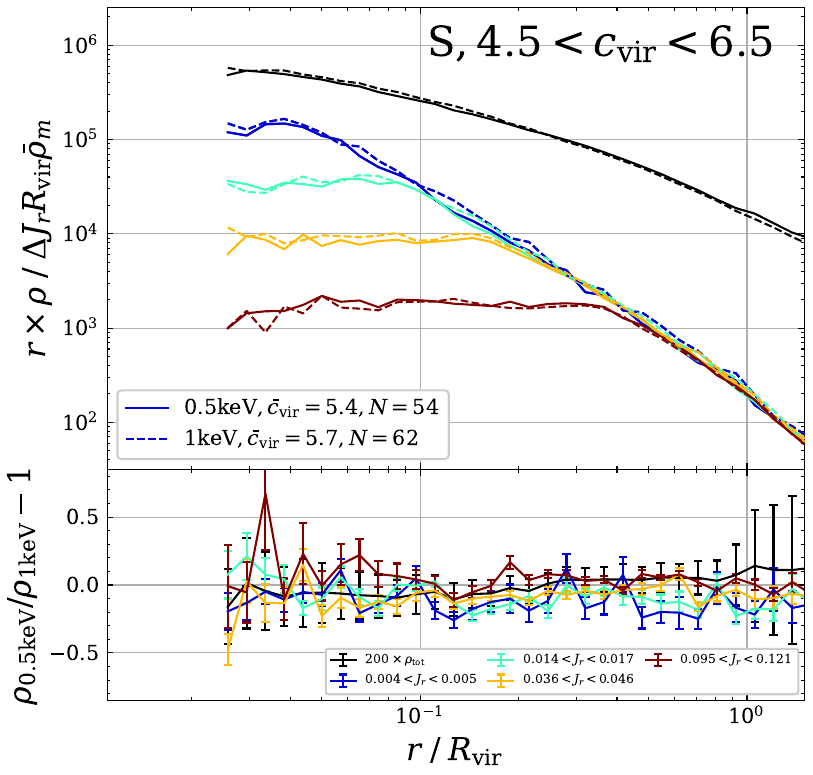}
        \end{minipage} &
        \begin{minipage}{\columnwidth}
        \centering            
        \includegraphics[scale=0.6]{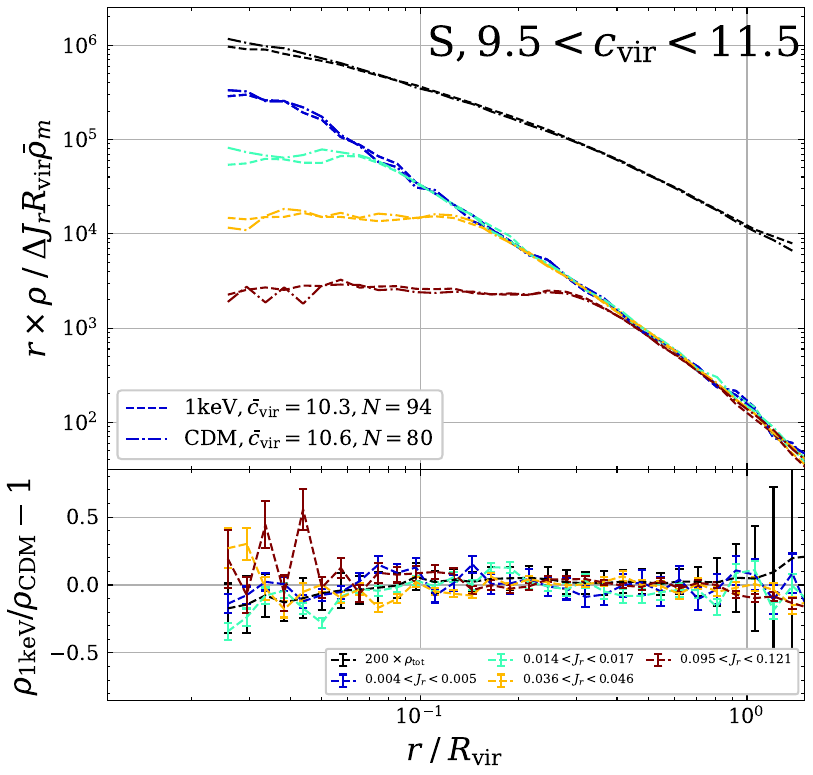}
        \end{minipage} 
    \end{tabular}
    \centering
    \caption{
    The comparison of the stacked density profiles of particles in halos in the mass range S and whose concentrations are in between $9.5<c_\mathrm{vir}<11.5$ (left) and $4.5<c_\mathrm{vir}<6.5$ (right).
    We compare $1$keV (dashed lines) and CDM (dot-dashed lines) cases in the left panel, and $0.5$keV (solid lines) and $1$keV (dot-dashed lines) cases in the right panel.
    $\bar{c}_\mathrm{vir}$ and $N$ shown in the legend indicate the mean value of $c_\mathrm{vir}$ and the number of halos in each subgroup.
    The bins for $J_r$ are the same as those in figure~\ref{fig:Jrdens_stacked_compar}.
    Bottom: the fractional difference between profiles shown in the top panel.
    Alt Text: The stacked density profiles of particles classified by radial action in CDM and WDM cases, showing that the difference in the profiles between dark matter models is attributed to the statistical difference in the concentration of halos across the models.
    }
    \label{fig:Jrdens_stacked_masscvir}
\end{figure*}
%%%%%%%%%%%%%%%%%%%%%%%%%%%%%%%%%%%%%%%%%%%%%%%%%%%%%%%%%%%%%%%%%%

% \newpage

%%--%%--%%--%%--%%--%%--%%--%%--%%--%%--%%--%%--%%--%%--%%--%%--%%
%%--%%--%%--%%--%%--%%--%%--%%--%%--%%--%%--%%--%%--%%--%%--%%--%%
\section{Discussion}\label{subsec:connection_Jr_p}
%%--%%--%%--%%--%%--%%--%%--%%--%%--%%--%%--%%--%%--%%--%%--%%--%%
%%--%%--%%--%%--%%--%%--%%--%%--%%--%%--%%--%%--%%--%%--%%--%%--%%

%%--%%--%%--%%--%%--%%--%%--%%--%%--%%--%%--%%--%%--%%--%%--%%--%%
%%--%%--%%--%%--%%--%%--%%--%%--%%--%%--%%--%%--%%--%%--%%--%%--%%
% \subsection{Connection between radial action and number of apocenter passages}\label{subsec:connection_Jr_p}
%%--%%--%%--%%--%%--%%--%%--%%--%%--%%--%%--%%--%%--%%--%%--%%--%%
%%--%%--%%--%%--%%--%%--%%--%%--%%--%%--%%--%%--%%--%%--%%--%%--%%

In the previous section, we see that the phase-space structure of halos exhibits the double power-law feature for the DM particles classified with the number of apocenter passages $p$ and the radial action $J_r$. In particular, their inner density profiles are found to commonly have a slope of $-1$ among the three DM models we analyzed. This feature is persistent even when dividing halo samples by the concentration parameters. This universal behavior suggests that there exists an intimate relation between $p$ and $J_r$, which remains unchanged irrespective of the DM model. 

\begin{figure*}
    \centering
    \includegraphics[width=16cm]{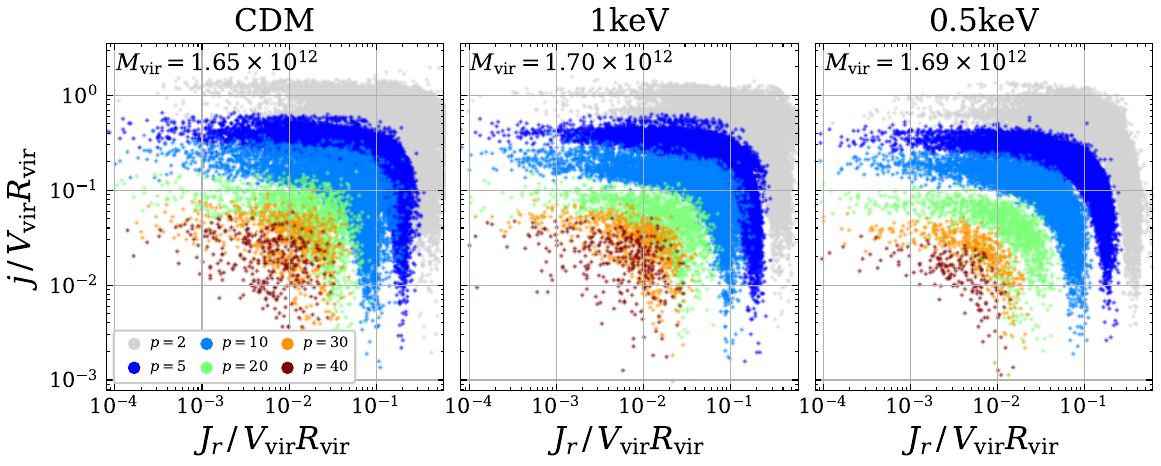}
    \caption{Radial action and angular momentum of particles in representative halos that are the same as those shown in the third raw in figures~\ref{fig:pdens_indiv} and \ref{fig:Jrdens_indiv}.
    The particles are color-coded corresponding to the number of apocenter passages as indicated in the legend.
    Alt Text: Relation between radial action and the number of apocenter passages, indicating that the inner region of $\rho(r;J_r)$ and $\rho(p;r)$ consist of mainly the same particles.
    }
    \label{fig:Jr_j_distri_example}
\end{figure*}

To investigate this relation, we indentify halos from three simulations that have nearly the same mass and position. Among these matched halos, we select one with a mass around $M_{\rm vir}\simeq1.7\times 10^{12}\,h^{-1}M_\odot$ as a representative example.
% which are identical among the DM models. 
We then plot the distribution of their DM particles for the number of apocenter passages $p=2$ (gray), $10$ (dark blue), $20$ (light blue), $30$ (orange), and $40$ (brown) in the plane of angular momentum $j$ and radial action $J_r$. The results are shown in figure \ref{fig:Jr_j_distri_example}, where the vertical and horizontal axes are normalized by $V_{\rm vir}\,R_{\rm vir}$. 

Interestingly, the distribution of the particles for a fixed $p$ has a characteristic shape, showing a rectangular pattern. That is, there are preferential values in $j$ and $J_r$, and DM particles are predominantly distributed with a scatter along either the preferential value of $J_r$ or $j$ up to a certain maximum value. Overall, irrespective of DM models, these preferential values systematically decrease as increasing $p$. 

From figure~\ref{fig:Jr_j_distri_example}, 
an important observation is that for a fixed $p$, particles having a small value of $j$ mostly lie around a certain value of $J_r$. Since the particles with small $j$ are likely to reside near the halo center and can contribute to the inner density profiles, this means that there is a tight link between $p$ and $J_r$ for the inner DM particles, which explains why the radial density profiles commonly show an inner slope of $-1$ for particles classified both with $p$ and $J_r$. 

On the other hand, particles residing in the outer part of halos can generally take various values of $j$, and their orbital radius tends to become larger as increasing $j$. However, as seen in figure~\ref{fig:Jr_j_distri_example}, there is an upper boundary in $j$ for particles with a fixed $p$. This implies that the density profiles for these particles sharply drop toward outer regions, with typically the slope of $-7\sim-9$. In contrast, particles with a fixed $J_r$ have various values of $p$, and  
their radial density profile is thus described by a superposition of those with different $p$. Hence, the outer profiles are smeared and have a shallower slope than those for the particles classified with $p$. These are all consistent with figures \ref{fig:pdens_compar} and \ref{fig:Jrdens_stacked_compar}. Nevertheless, the origin of the outer slope of $-3.5$ for the particles classified by $J_r$ still remains puzzling, together with the common inner slope of $-1$, and we need to look for other dynamical features imprinted in the halo formation and accretion history.

%%%%%%%%%%%%%%%%%%%%%%%%%%%%%%%%%%%%%%%%%%%%%%%%%%%%%%%%%%%%%%%%%%
\section{Conclusions and outlook}\label{sec:conclusion}
%%--%%--%%--%%--%%--%%--%%--%%--%%--%%--%%--%%--%%--%%--%%--%%--%%
%%--%%--%%--%%--%%--%%--%%--%%--%%--%%--%%--%%--%%--%%--%%--%%--%%

In this paper, we have characterized the inner structure of dark matter halos, paying a particular attention to their phase-space properties. Specifically, making use of the dynamical information on dark matter (DM) particles, 
we have quantified the radial density structures in two complementary ways. One is to classify the DM particles by the number of apocenter passages $p$, following our previous works \citep{Sugiura2020, Enomoto2023, Enomoto2024}. In doing so, we have kept track of the particle trajectories around the halo and count the number $p$ from the $1,000$ snapshots in $0\leq z\leq 5$. Another is to compute the radial action $J_r$ for each DM particle, and to characterize the particle distribution with $J_r$, based on the method presented in \citet{Pontzen2013}. With these characterizations, we studied the properties of radial density profiles not only in the CDM model but also in the WDM models with different DM masses, $m_{\rm DM}=0.5$ and $1$\,keV, both of which are observationally disfavared but are expected to exhibit distinct features compared to the CDM model. 

Our important findings are summarized below:

\begin{description}
    \item[(i)] For all of the three DM models, the density profiles exhibit double power-law features for particles classified not only by $p$ but also by $J_r$, with shallow inner slope and steep outer slope. These common behaviors are associated to the tight relation between $p$ and $J_r$ seen in figure~\ref{fig:Jr_j_distri_example}.  
    \item [(ii)] In both classifications, the inner profiles commonly have the slope of $-1$. On the other hand, the outer profiles differ for the two classifications. While the particles classified by $p$ have a rather steep slope of $-6 \sim -8$, those classified by $J_r$ exhibits a common slope of $-3.5$. 
    \item[(iii)] Among the three DM models, there are systematic differences in the amplitude and characteristic scale of the radial profiles. As increasing the mass of DM, the inner amplitude becomes larger, while the outer amplitude decreases, leading to a reduced characteristic scale. 
   These trends are, in fact, attributed to differences in halo concentration. We found that for halos with the same concentration parameter, the density profiles for particles classified by $J_r$ are nearly identical across different DM models.
    \end{description}
 
% \ATcom{As discussed in \cite{Enomoto2023,Enomoto2024}, summing up the individual contribution, double power-law structures found in the particle distribution classified by $p$ can recover the total density profile. Together with well-calibrated characteristic density and scale of the double power-law profile, the predicted total profile is shown to be comparable to or even better than the Einasto profile \citep{einasto1965construction}. One anticipates that this must also hold for the double power-law features of the particles classified by $J_r$. In other words, the origin of the universal features of radial halo profiles, long debated in the literature, may be encapsulated in the double power-law profiles analyzed in this paper. In particular, the shallow cusp of the total halo profile is likely linked to 
% the inner slope of $-1$ seen in both classification schemes. }
As discussed in \citet{Enomoto2023} and \citet{Enomoto2024}, by summing the individual contributions, the double power-law structures found in particle distributions classified by $p$ can accurately reproduce the total density profile. With a well-calibrated characteristic density and scale for the double power-law profile, the predicted total profile has been shown to be comparable to, or even better than, the Einasto profile \citep{einasto1965construction}. It is expected that this result should also apply to the double power-law features of particles classified by $J_r$. In other words, the origin of the universal features in radial halo profiles, a long-debated topic in the literature, may be captured by the double power-law profiles analyzed in this paper. Specifically, the shallow cusp of the total halo profile is likely related to the inner slope of $-1$ observed in both classification schemes.

Although we do not have a clear physical explanation of the double power-law feature yet, the result  that this characteristic persists regardless of particle classification provides an important hint. That is, one can study the radial phase-space structure not only with the number of apocenter passages $p$ but also via the radial action $J_r$, which is better suited for characterizing a relaxed halo system and has been widely used in the literature.

Following previous studies (e.g., \cite{Pontzen2013, Muni2024MNRAS}), we express the enclosed mass for particles classified by $J_r$ at radius $r$, denoted by $M(<r;,J_r)$, as follows: 
%%%%%%%%%%%%%%%%%%%%%%%%%%%%%%%%%%%%%%%%%%%%%%
\begin{eqnarray}\label{eq:innermass}
    M(<r;J_r) = m_p\int_0^{\infty} p(j,J_r) P(<r;j,J_r) dj,
\end{eqnarray}
%%%%%%%%%%%%%%%%%%%%%%%%%%%%%%%%%%%%%%%%%%%%%%
where $m_p$ is the mass of the particle in simulation. 
$p(j, J_r)$ represents the joint distribution function of particles with radial action $J_r$ and angular momentum $j$\footnote{The joint distribution function $p(j,J_r)$ here is computed by integrating the full distribution function of action variables $f(\boldsymbol{J} \equiv (J_r,j,j_z))$ over $j_z$ (see equation (B2) in \citet{Pontzen2013}).
As a theoretical model for $f(\boldsymbol{J})$, functions proposed in \citet{Posti2015MNRAS} and \citet{Williams2015MNRAS} are widely used to model our galaxy (e.g., \cite{Hattori2021MNRAS, Binney2023MNRAS, Binney2024MNRAS}) or M31-like galaxy \citep{Gherghinescu2024MNRAS}.}.
The function  $P(<r;j,J_r)$ represents the fraction of time that a particle with orbital parameters $(j,J_r)$ spends within $r$, which is defined by 
%%%%%%%%%%%%%%%%%%%%%%%%%%%%%%%%%%%%%%%%%%%%%%
\begin{eqnarray}
    P(<r;j,J_r) = \frac{2}{t_\mathrm{orb}(j,J_r)} \int_{r_\mathrm{peri}}^r \frac{dr}{v_r(r)},
\end{eqnarray}
%%%%%%%%%%%%%%%%%%%%%%%%%%%%%%%%%%%%%%%%%%%%%%
where $t_\mathrm{orb}(j,J_r)$ is the orbital period of the particle.
With the expression given above, 
the density profile for the particles for a given $J_r$, $\rho(r;\,J_r)$,  is given by
%%%%%%%%%%%%%%%%%%%%%%%%%%%%%%%%%%%%%%%%%%%%%%
\begin{equation}
    \rho(r;J_r) = \frac{1}{4\pi r^2}\frac{dM(<r;J_r)}{dr}
    \propto r^{-3}M(<r;J_r).
\end{equation}
%%%%%%%%%%%%%%%%%%%%%%%%%%%%%%%%%%%%%%%%%%%%%%
Equation~\eqref{eq:innermass} says that the double-power-law nature of $\rho(r;J_r)$ is encapsulated in the functional form of $p(j,J_r)$ and $P(<r;j,\,J_r)$.  Several studies have provided theoretical predictions for them and there are some arguments on how the power-law nature appears\footnote{For example, in a nearly flat potential, $P(<r;j,J_r)$ can be approximated as $P(<r;j,J_r) \propto r/v_r(r) \sim r$ \citep{Dalal2010, Lithwick2011ApJ}. 
If $p(j,J_r)$ follows a power law, such that $p(j,J_r) \propto j^{\alpha}$, then the slope of $\rho(r;J_r)$ behaves as $\rho(r;J_r) \propto r^{\alpha-1}$.}. 
Quantifying their functional forms using $N$ N-body simulations is also straightforward, and the analysis is currently in progress. 
We will report the results in the near future.

\if0
In our future work, we will investigate the shape of $P(<r;j,J_r)$ and explore the connection between $\rho(r;J_r)$ and $p(j,J_r)$. 
Additionally, we aim to determine whether the asymptotic slope of $\rho(r;J_r)$ approaches $-1$, assuming a simple form for $p(j,J_r)$
\footnote{ \citet{Pontzen2013} showed from maximum entropy theory that the shape of $p(j,J_r)$ behaves asymptotically like $p(j,J_r) \propto j$ in the $j\to 0$ limit (see equation~B2 therein).
Generally, the shape of the angular momentum distribution reflects the available states of the angular momentum vector. 
If particles with the same energy have isotropic velocity distribution, their angular momentum distribution is proportional to $j$ in $j \to 0$ limit (\cite{Lu2006}, see equation~16 therein).}.
\fi

% Suppose one can produce a theoretical prediction for  $p(j,J_r)$ by extending previous works (e.g., \cite{Pontzen2013, Burger2021}). 
% In that case, the double-power law nature might be explained from the first principle, which in turn may explain the origin of the universality of the density profiles and pseudo-density profiles.

From an observational perspective, the phase-space information of stellar halos could serve as a tool to validate the universal features we have discovered. With the data provided by the Gaia satellite \citep{Gaia2016A&A}, along with complementary spectroscopic surveys such as GALAH \citep{DeSilva2015MNRASGALAH}, APOGEE \citep{Majewski2017APOGEE} and H3 \citep{Conroy2019ApJ}, six-dimensional phase-space information for more than a billion stars is now available (\cite{Gaia2021EDR3}). This wealth of data has been leveraged to explore the formation history of the Milky Way (e.g., \cite{Naidu2020ApJ, Belokurov2023MNRAS, Donlon2024MNRAS}; see also recent reviews by \cite{Helmi2020review, Deason2024review, Bonaca2024review}).
Recent studies based on cosmological hydrodynamical simulations have reported a correlation between the kinematics of dark matter and stellar halos \citep{Herzog-Arbeitman2018PhRvL, Necib2019, Deason2020MNRAS, Genina2023}. This correlation has been used to estimate the velocity distribution of dark matter around the Sun by leveraging stellar data \citep{Herzog-Arbeitman2018JCAP, Necib2019Gaia}. Notably, the radial density profile of the stellar halo in our galaxy exhibits a broken power law (e.g., \cite{Watkins2009MNRAS, Deason2011MNRAS, Sesar2011ApJ, Pila-Diez2015A&A, Han2022AJ}).
Therefore, if we classify stars in the stellar halo by their radial action, we may uncover a similar universal feature in their density profiles, analogous to that observed for dark matter particles. If this proves to be the case, the slope, amplitude, and scale radius of these density profiles could potentially serve as key indicators for distinguishing between different dark matter models.

However, several challenges must be addressed to observationally verify dark matter models. 
First, it is crucial to clarify the conditions and extent to which the distributions of dark matter and stars, classified by radial action, align using cosmological hydrodynamical simulations. 
Fortunately, radial action can be calculated from a single snapshot, unlike the number of apocenter passages, making it potentially less challenging to extend our results from $N$-body simulations to hydrodynamical simulations. 
If the density profiles of stars classified by the radial action resemble those of dark matter, they could serve as powerful new tools for distinguishing between dark matter models or, at the very least, for verifying the CDM model.
Additionally, investigating the impact of baryonic processes, including AGN and supernova feedback, on our findings would be highly insightful.
It would be practically important to provide a robust prediction of the phase-space structure of halos that could be a basis for the observational verification of the dark matter model.

Second, to develop a robust theoretical prediction, we must analyze simulations that include a larger number of halos (e.g., \cite{Ishiyama2021MNRASUchuu}). As shown in Table~\ref{tab:halo_catalog}, each mass range contains only a few hundred halos at most, which may not provide a sufficiently reliable prediction for the average density profile across each mass range. 
Larger simulations with higher resolution are necessary to generate theoretical predictions robust enough for comparison with observational data.

It is also worthwhile to investigate simulations of warm dark matter (WDM) within the dark matter mass ranges allowed by observations, as well as alternative dark matter models such as self-interacting dark matter (SIDM) and fuzzy dark matter (FDM). If we can robustly examine the dependence of the amplitude or scale of $\rho(r;J_r)$ on cosmology or dark matter models, these results could offer viable predictions for comparison with observational data.

\begin{ack}
This work was
supported in part by MEXT/JSPS KAKENHI Grant
Number JP20H05861, JP21H01081
(AT and TN), JP22K03634, JP24H00215 and JP24H00221 (TN). 
% We also acknowledge financial support from Japan Science and Technology Agency (JST) AIP Acceleration Research Grant Number JP20317829 (AT and TN). 
% YE is also supported by JST, the establishment of university fellowships towards the creation of science and technology innovation, Grant Number JPMJFS2123, and
YE is also supported by JSPS KAKENHI Grant Number 24KJ1361.
Numerical computations were carried out at Yukawa Institute Computer Facility, and Cray XC50 at the Center for Computational Astrophysics, National Astronomical Observatory of Japan. 
This research was supported by MEXT as “Program for Promoting Researches on the Supercomputer Fugaku” (Multi-wavelength Cosmological Simulations for Next-generation Surveys, JPMXP1020230407, project ID hp230202) and used computational resources of supercomputer Fugaku provided by the RIKEN Center for Computational Science.
Finally, YE gratefully thanks TAP colleagues for their warm support and valuable discussions.

\end{ack}

%%--%%--%%--%%--%%--%%--%%--%%--%%--%%--%%--%%--%%--%%--%%--%%--%%
%%--%%--%%--%%--%%--%%--%%--%%--%%--%%--%%--%%--%%--%%--%%--%%--%%
\appendix 
\section{Convergence tests of the methods evaluating $J_r$}\label{sec:app1}
%%--%%--%%--%%--%%--%%--%%--%%--%%--%%--%%--%%--%%--%%--%%--%%--%%
%%--%%--%%--%%--%%--%%--%%--%%--%%--%%--%%--%%--%%--%%--%%--%%--%%

%\section*{Case of single paragraph}

%%%%%%%%%%%%%%%%%%%%%%%%%%%%%%%%%%%%%%%%%%%%%%%%%%%%%%%%%%%%%%%%%%
\begin{figure*}
    \centering
    \includegraphics[width=16cm]{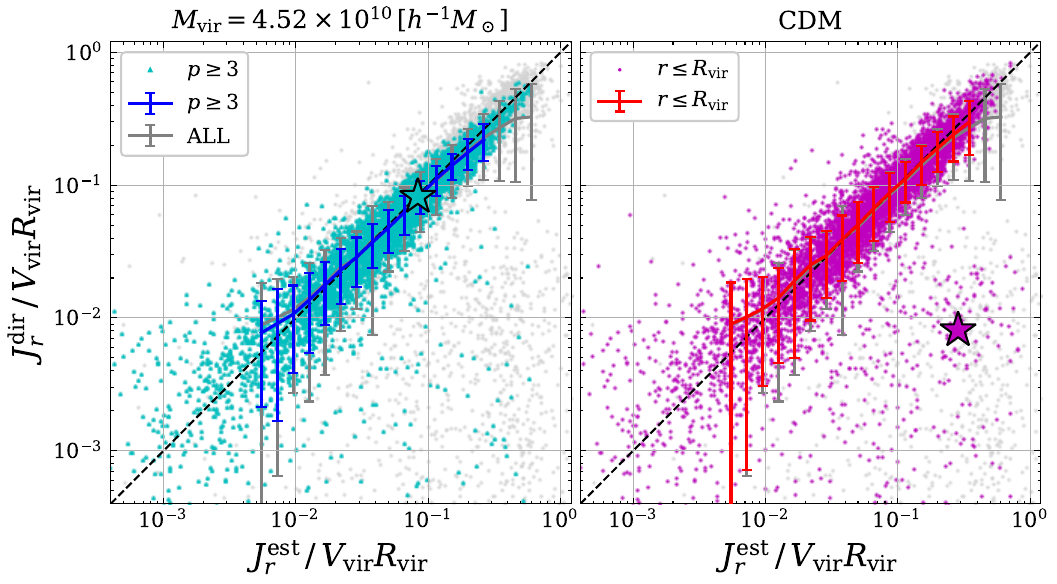}
    \caption{Comparison between $J_r^\mathrm{dir}$ and $J_r^\mathrm{est}$ of particles in a representative CDM halo.
    In both panels, we show all the particles within $2.5R_\mathrm{vir}$ as thin grey markers and their mean and RMS scatter as grey solid lines with error bars.
    Left: particles with $p\geq 3$ (cyan markers) and their RMS scatter (blue error bars) are shown.
    Right: particles located within $1R_\mathrm{vir}$ from the center of the halo (magenta markers) and their RMS scatter (magenta error bars) are also shown.
    In both cases, error bars are evaluated in bins with more than $100$ particles.
    The stars correspond to the particles whose trajectories are shown in figure~\ref{fig:Jr_est_to_dir_traject}.
    Alt Text: Convergence test between two methods computing radial actions $J_r$, showing that both methods provide values of $J_r$ that coincide with each other in general.
    }
    \label{fig:Jr_est_to_dir_distri}
\end{figure*}
%%%%%%%%%%%%%%%%%%%%%%%%%%%%%%%%%%%%%%%%%%%%%%%%%%%%%%%%%%%%%%%%%%

%%%%%%%%%%%%%%%%%%%%%%%%%%%%%%%%%%%%%%%%%%%%%%%%%%%%%%%%%%%%%%%%%%
\begin{figure}
    \centering
    \includegraphics[width=8cm]{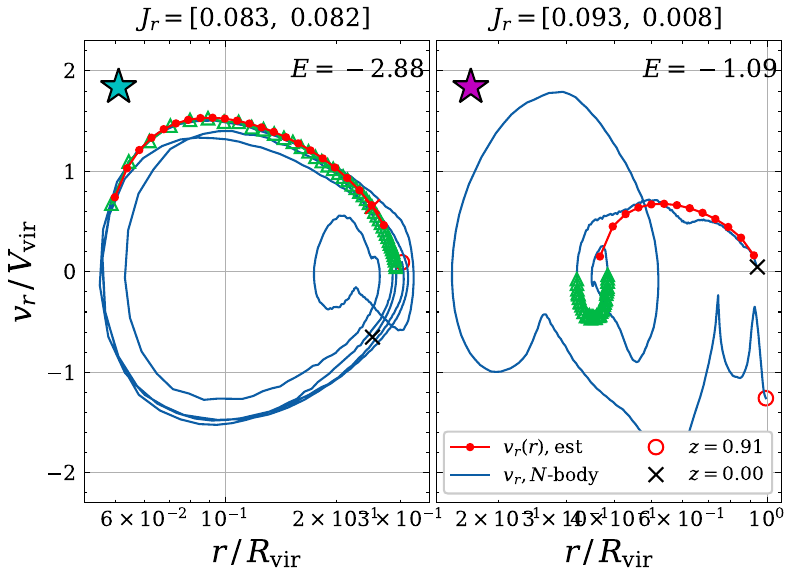}
    \caption{
    The phase-space trajectories of particles indicated as stars in figure~\ref{fig:Jr_est_to_dir_distri}.
    Both panels show the particles' trajectory from $z=0.91$ (red circle) to $z=0$ (black cross) as blue solid lines.
    Note there are 600 snapshots between the redshifts.
    The red markers indicate the trajectory reconstructed from the energy and angular momentum evaluated at $z=0$ (black cross, or red cross which has mirrored radial velocity of the black cross), and $J_r^\mathrm{est}$ is evaluated from the area enclosed by the line connecting the markers and x-axis ($v_r=0$).
    $J_r$ shown at the title of both panels indicate the values of each $[J_r^\mathrm{est},J_r^\mathrm{dir}]$ in units of $R_\mathrm{vir}V_\mathrm{vir}$, and the energies in units of $M_\mathrm{vir}V^2_\mathrm{vir}$ are shown at the top right corner. 
    Left: a representative particle shown in the left panel of figure~\ref{fig:Jr_est_to_dir_distri} (cyan star) that is $p=10$ and whose $J_r^\mathrm{est}$ and $J_r^\mathrm{dir}$ matches well.
    Right: a representative particle shown in the right panel of figure~\ref{fig:Jr_est_to_dir_distri} (magenta star) that is located inside $1R_\mathrm{vir}$ and whose $J_r^\mathrm{est}$ and $J_r^\mathrm{dir}$ differ significantly.
    Alt Text: The representative trajectories of particles that have a well-defined value of its radial action (left panel) and not (right panel).
    }
    \label{fig:Jr_est_to_dir_traject}
\end{figure}
%%%%%%%%%%%%%%%%%%%%%%%%%%%%%%%%%%%%%%%%%%%%%%%%%%%%%%%%%%%%%%%%%%

%%%%%%%%%%%%%%%%%%%%%%%%%%%%%%%%%%%%%%%%%%%%%%%%%%%%%%%%%%%%%%%%%%
\begin{figure*}
    \begin{tabular}{ccc}
        \begin{minipage}{0.66\columnwidth}
        \centering            \includegraphics[scale=0.56]{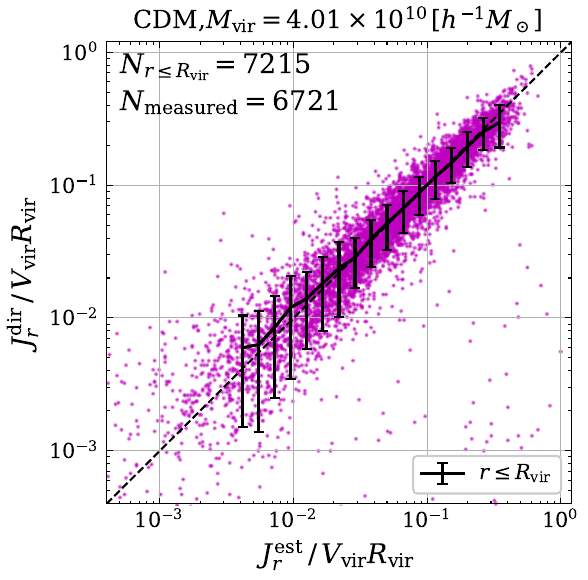}
        \end{minipage} &
        \begin{minipage}{0.66\columnwidth}
        \centering
        \includegraphics[scale=0.56]{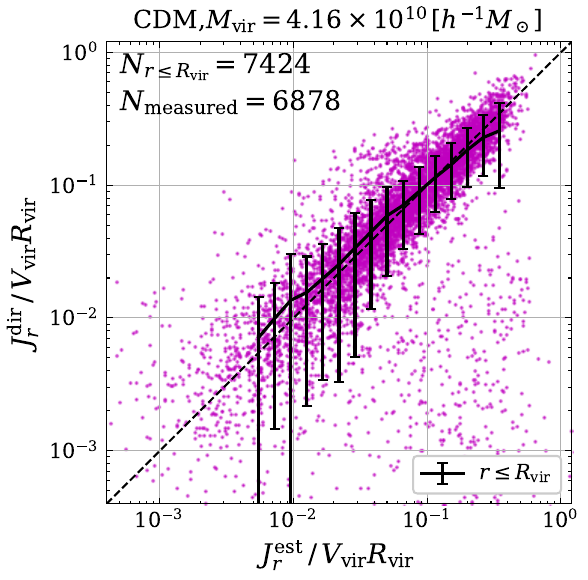}
        \end{minipage} &
        \begin{minipage}{0.66\columnwidth}
        \centering            \includegraphics[scale=0.56]{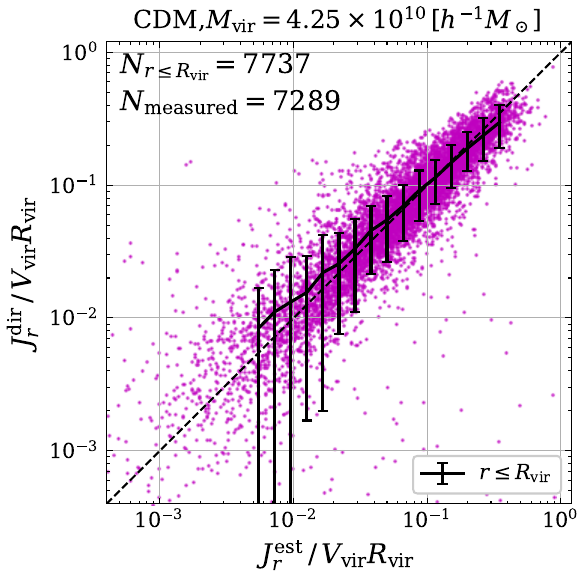}
        \end{minipage} \\
        \begin{minipage}{0.66\columnwidth}
        \centering            \includegraphics[scale=0.56]{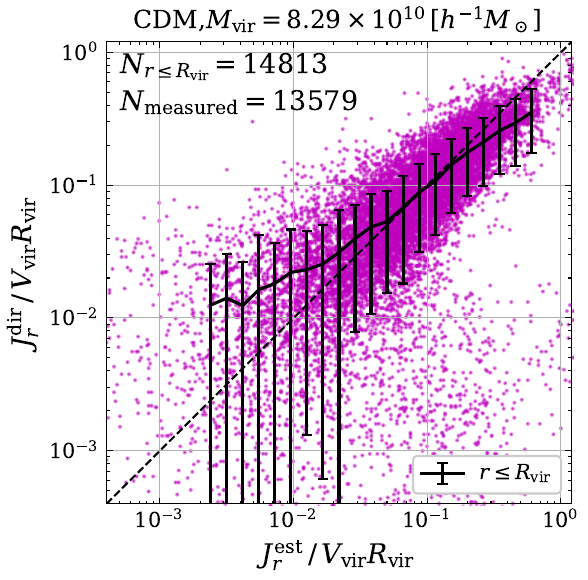}
        \end{minipage} &
        \begin{minipage}{0.66\columnwidth}
        \centering            \includegraphics[scale=0.56]{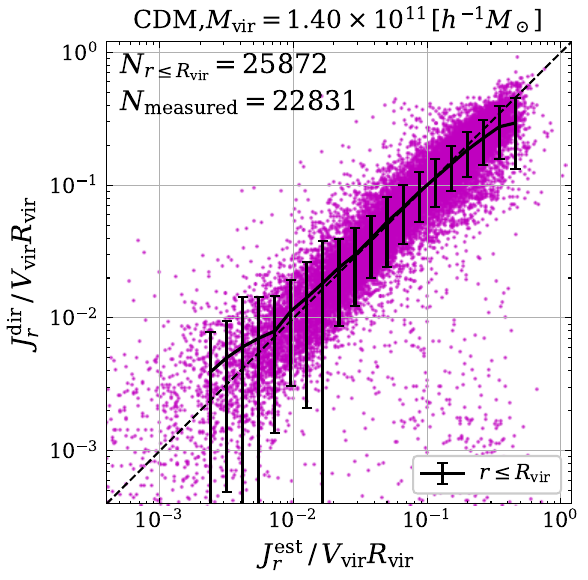}
        \end{minipage} &
        \begin{minipage}{0.66\columnwidth}
        \centering            \includegraphics[scale=0.56]{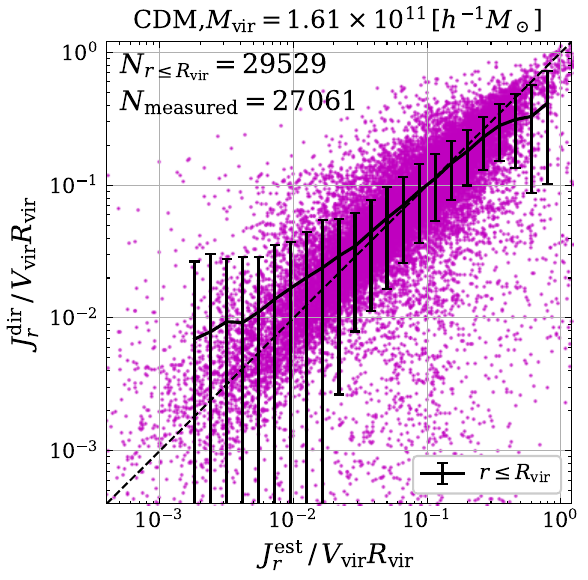}
        \end{minipage} \\
        \begin{minipage}{0.66\columnwidth}
        \centering            \includegraphics[scale=0.56]{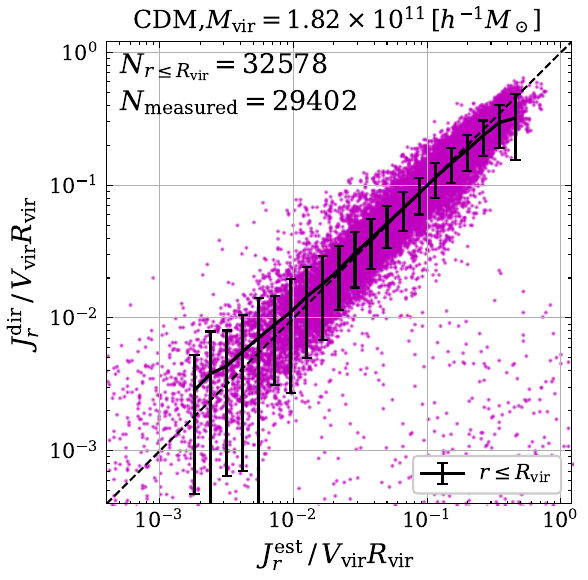}
        \end{minipage} &
        \begin{minipage}{0.66\columnwidth}
        \centering            \includegraphics[scale=0.56]{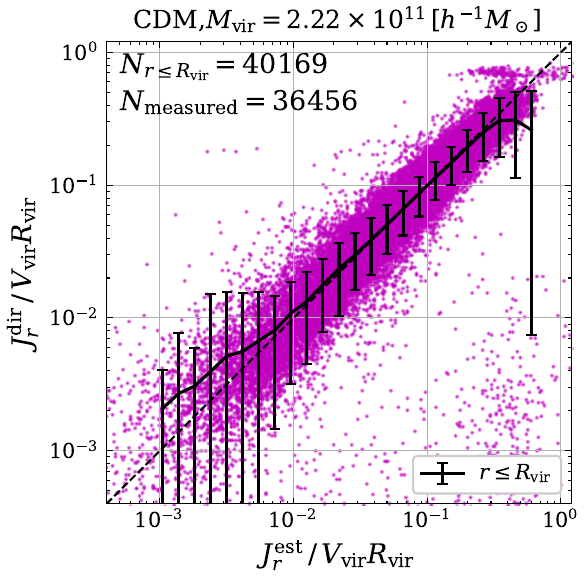}
        \end{minipage} &
        \begin{minipage}{0.66\columnwidth}
        \centering            \includegraphics[scale=0.56]{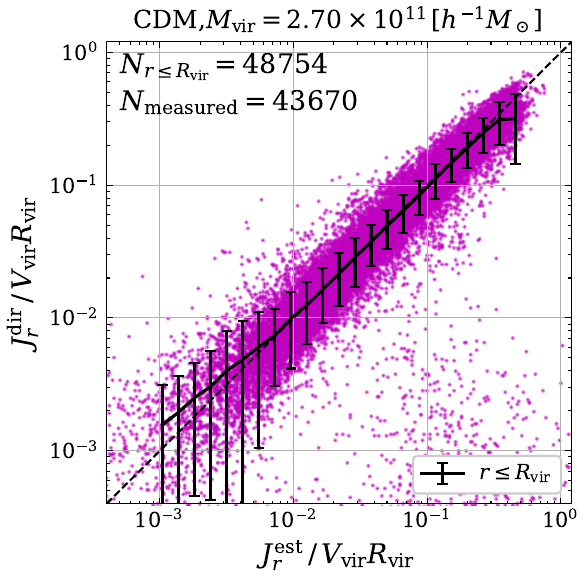}
        \end{minipage}
    \end{tabular}
    \caption{Comparison between $J_r^\mathrm{dir}$ and $J_r^\mathrm{est}$ of particles in a representative CDM halos.
    In each panel, we show all the particles within $1R_\mathrm{vir}$ from the center of the halo as magenta markers, and their mean and RMS scatter as black solid lines with error bars.
    Error bars are evaluated in bins where there are more than $100$ particles.
    $N_{r\leq R_\mathrm{vir}}$ and $N_\mathrm{measured}$ indicate the number of particles that locate within the virial radius $R_\mathrm{vir}$ and particles whose $J_r^\mathrm{dir}$ and $J_r^\mathrm{est}$ can be measured respectively.
    $J_r^\mathrm{dir}$ can be measured for roughly $90\%$ of particles inside the virial radius.
    Alt Text: Convergence test of the methods for calculating radial action in nine representative CDM halos, demonstrating that both approaches generally yield consistent values of $J_r$.
    }
    \label{fig:Jr_est_to_dir_distri_halos}
\end{figure*}
%%%%%%%%%%%%%%%%%%%%%%%%%%%%%%%%%%%%%%%%%%%%%%%%%%%%%%%%%%%%%%%%%%

%%%%%%%%%%%%%%%%%%%%%%%%%%%%%%%%%%%%%%%%%%%%%%%%%%%%%%%%%%%%%%%%%%
\begin{figure}
    \centering
    \includegraphics[width=8cm]{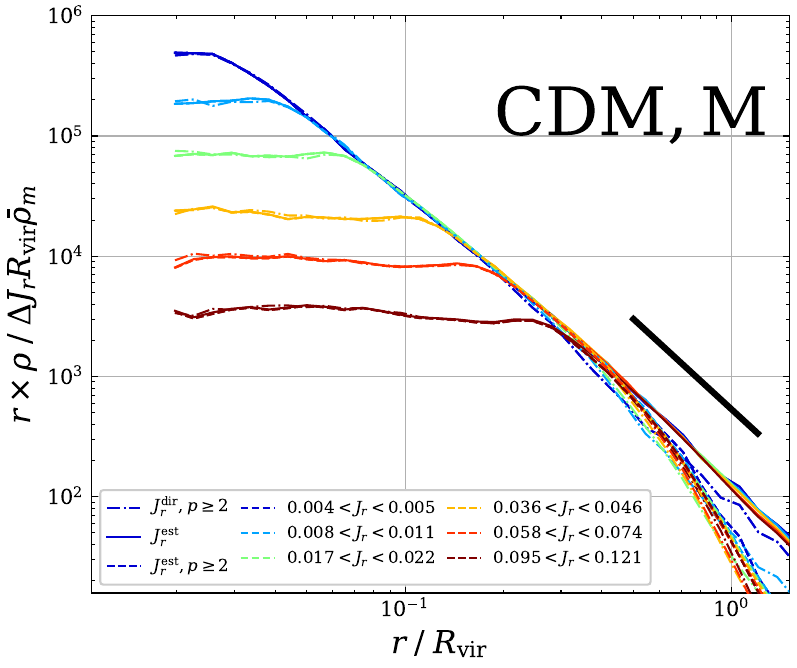}
    \caption{The stacked density profiles of particles classified by radial actions $J_r^\mathrm{est}$ and $J_r^\mathrm{dir}$ in CDM halos in the mass range M.
    The results from the classification by $J_r^\mathrm{est}$ (solid lines), and those of particles with $p\geq 2$ classified by $J_r^\mathrm{est}$ (dashed lines) and $J_r^\mathrm{dir}$ (dot-dashed lines) are shown.
    Alt Text: Convergence test of the methods for calculating radial action, indicating that the choice of the methods does not affect our findings in section~\ref{section:results}.
    }
    \label{fig:Jrdens_dir_est_CDM}
\end{figure}
%%%%%%%%%%%%%%%%%%%%%%%%%%%%%%%%%%%%%%%%%%%%%%%%%%%%%%%%%%%%%%%%%%

In this appendix, we demonstrate that the choice of the method for evaluating $J_r$ does not affect our results.
As introduced in section~\ref{subsec:count_Jr}, we utilize the method of integrating equation~\ref{eq:def_of_jr} to calculate the radial action of each particle.
However, this method assumes that particles' energy $E$ or angular momentum $j$ are conserved while the particles complete their orbit.
In order to assess the validity of this, we utilize another method that directly evaluates equation~\ref{eq:def_of_jr} using the radial velocity of particles measured at each snapshot and compare $J_r$ computed by both methods.
In the latter method, we monitor the radial velocities of particles between the latest passage of their pericenter and apocenter and substitute them into equation~\ref{eq:def_of_jr}.
Therefore, this method can be applied to particles that have passed their apocenter at least once (i.e., $p=1$) but not to particles that are currently orbiting between apoapsis for the first time ever though they are bounded particles.
The former method can adopt all the bounded particles, and this difference affects the outer region of the stacked density profiles as will be shown in figure~\ref{fig:Jrdens_dir_est_CDM}.

In figure~\ref{fig:Jr_est_to_dir_distri}, we compare the radial action of particles in a representative CDM halo evaluated by the methods introduced above.
Here we refer to $J_r$ calculated with the method introduced in section~\ref{subsec:count_Jr} as $J_r^\mathrm{est}$ and that calculated from the integral of measured radial velocity as $J_r^\mathrm{dir}$.
In both panels, we show all the particles within $2.5R_\mathrm{vir}$ from the center of the halo as thin grey markers, and its mean and RMS scatter are shown as grey solid lines with error bars.
We also plot the values of particles with $p\geq3$ (cyan markers in the left panel), and those residing within $1R_\mathrm{vir}$ from the center of the halo (magenta markers in the right panel), and their scatter as errorbars with the corresponding color.
$J_r^\mathrm{est}$ and $J_r^\mathrm{dir}$ match well as the mean values indicate, and the scatter is smaller for particles with $p\geq3$ and those residing within $1R_\mathrm{vir}$.
Therefore, the radial actions match better for bounded particles that have a large number of $p$ or are orbiting within the virial radius.

However, plenty of particles have lower values of $J_r^\mathrm{dir}$ compared with their $J_r^\mathrm{est}$ (see bottom right in both panels of figure~\ref{fig:Jr_est_to_dir_distri}), although the number of such particles decreases significantly after setting $p\geq 3$ or $r\leq R_\mathrm{vir}$ thresholds.
In order to clarify the characteristics of these particles, in figure~\ref{fig:Jr_est_to_dir_traject}, we show phase-space trajectories of two representative particles that are shown as cyan and magenta stars in figure~\ref{fig:Jr_est_to_dir_distri}.
The left particle has $p=10$ and its $J_r^\mathrm{est}$ and $J_r^\mathrm{dir}$ are almost identical, while the right one resides within $1R_\mathrm{vir}$ but its radial action quite differs from each other.
In both panels, the red dots connected by lines depict trajectories reconstructed from the measured potential, energy, and angular momentum evaluated at $z=0$ (black cross), and their $J_r^\mathrm{est}$ are evaluated from the area enclosed by the lines and the x-axis ($v_r=0$).
The green triangles depict the measured trajectories of the latest orbit, which is used for evaluating $J_r^\mathrm{dir}$ as the area enclosed by the line connecting the triangles and the x-axis.
In the left panel, the particle has a periodic orbit and smaller orbital period, and its reconstructed trajectory (red dots) and measured one (green triangles) match well, resulting in the well-matched values of the radial action.
On the other hand, the particle shown in the right panel has a longer orbital period with large energy and fluctuates around $v_r=0$.
Because of irregular fluctuation due to substructures, many particles with long orbital periods tend to have small values of $J_r^\mathrm{dir}$ while radial actions of particles residing deep inside the virial radius are not affected by the perturbation

% \footnote{In principle, radial action of such particles are conserved because they are adiabatic invariant. \citet{Burger2021} showed the radial action of particles with long orbital periods oscillates larger due to the time evolution of potential compared with those with shorter orbital periods.} 

In figure~\ref{fig:Jr_est_to_dir_distri_halos}, we also show $J_r^\mathrm{est}$ and $J_r^\mathrm{dir}$ of particles inside the virial radius for other nine representative halos.
As the mean values indicate (black lines), most halos have well-matched values of radial action, and some particles have smaller values of $J_r^\mathrm{dir}$ compared with their $J_r^\mathrm{est}$ due to the reason discussed above.

Finally, to demonstrate that our results of stacked profiles are not sensitive to the choice of methods evaluating radial action, we show the stacked density profiles of particles classified by $J_r^\mathrm{est}$ (solid lines) in figure~\ref{fig:Jrdens_dir_est_CDM}.
We also show the stacked profiles of particles with $p\geq 2$ classified by $J_r^\mathrm{est}$ (dashed lines), and $J_r^\mathrm{dir}$ (dot-dashed lines).
For the radii below $0.5R_\mathrm{vir}$, every profile for the same value of radial action matches each other while they vary for $r\geq 0.5R_\mathrm{vir}$.
Since we can not measure $J_r^\mathrm{dir}$ for particles that have not passed their apocenter until $z=0$ (i.e., $p=1$) because their trajectory has not closed, density profiles classified by $J_r^\mathrm{dir}$ are lowered.

However, this only affects the outer region of density profiles and it does not change our results below $0.5R_\mathrm{vir}$ even if we select particles that are $p\geq 2$ (see dot-dashed and solid lines in figure~\ref{fig:Jrdens_dir_est_CDM}).
We should note that the unmatch between the profiles for $r\geq 0.5R_\mathrm{vir}$ does not mean our results in these radii are not robust.
The profiles of particles classified by $J_r^\mathrm{dir}$ are lowered just because we can not compute $J_r^\mathrm{dir}$ for particles that have not passed their apocenter yet, while we can define radial action $J_r^\mathrm{est}$ for those particles.
We should note that it is inherently difficult to evaluate radial action for particles that infall to halo for the first time because they usually have long orbital periods and significantly oscillate due to substructures (see \cite{Burger2021}).

%\section{Case of two or more paragraphs}

%%--%%--%%--%%--%%--%%--%%--%%--%%--%%--%%--%%--%%--%%--%%--%%--%%
%%--%%--%%--%%--%%--%%--%%--%%--%%--%%--%%--%%--%%--%%--%%--%%--%%
\section{Convergence test against the resolution of $N$-body simulation}\label{sec:app2}
%%--%%--%%--%%--%%--%%--%%--%%--%%--%%--%%--%%--%%--%%--%%--%%--%%
%%--%%--%%--%%--%%--%%--%%--%%--%%--%%--%%--%%--%%--%%--%%--%%--%%

%%%%%%%%%%%%%%%%%%%%%%%%%%%%%%%%%%%%%%%%%%%%%%%%%%%%%%%%%%%%%%%%%%
\begin{figure*}
    \begin{tabular}{ccc}
        \begin{minipage}{0.66\columnwidth}
        \centering            \includegraphics[scale=0.56]{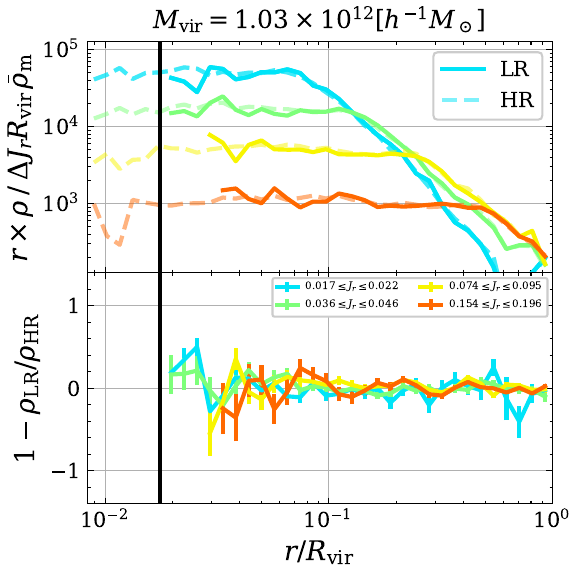}
        \end{minipage} &
        \begin{minipage}{0.66\columnwidth}
        \centering
        \includegraphics[scale=0.56]{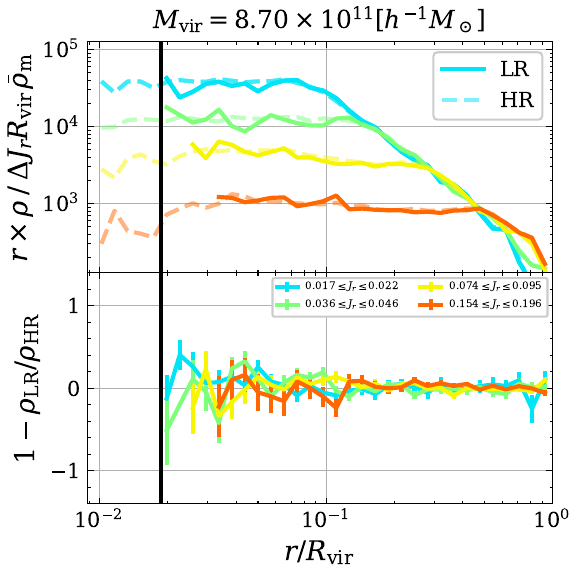}
        \end{minipage} &
        \begin{minipage}{0.66\columnwidth}
        \centering            \includegraphics[scale=0.56]{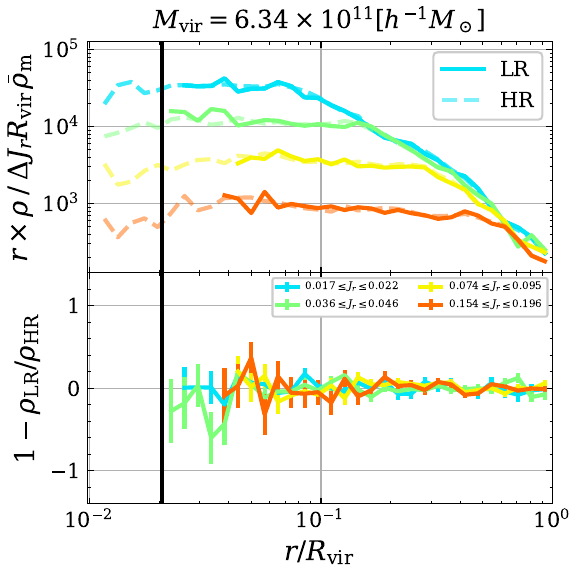}
        \end{minipage} \\
        \begin{minipage}{0.66\columnwidth}
        \centering            \includegraphics[scale=0.56]{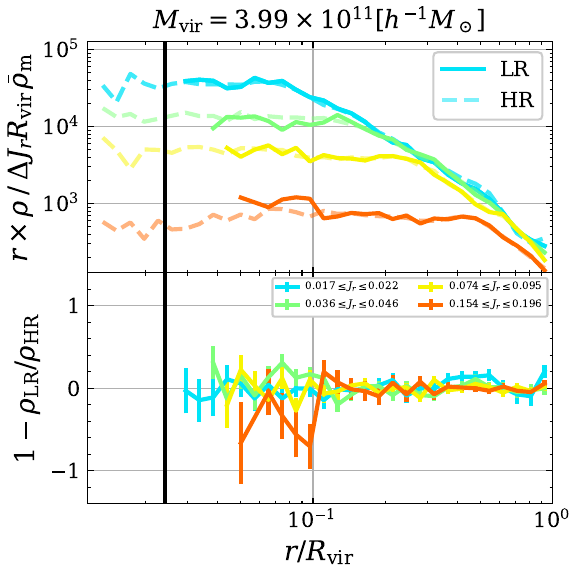}
        \end{minipage} &
        \begin{minipage}{0.66\columnwidth}
        \centering            \includegraphics[scale=0.56]{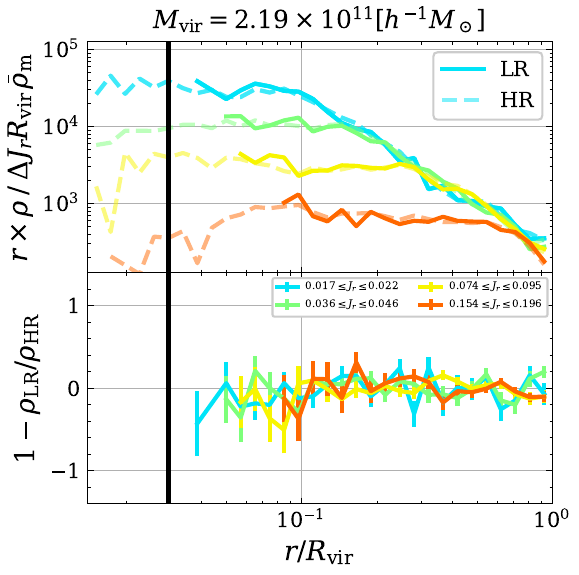}
        \end{minipage} &
        \begin{minipage}{0.66\columnwidth}
        \centering            \includegraphics[scale=0.56]{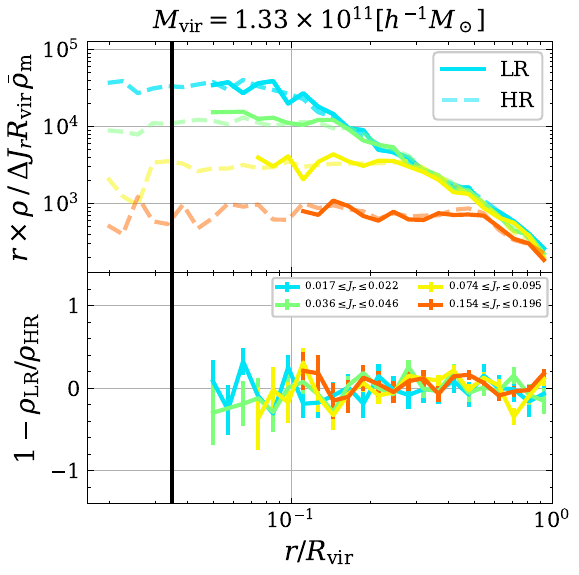}
        \end{minipage} \\
        \begin{minipage}{0.66\columnwidth}
        \centering            \includegraphics[scale=0.56]{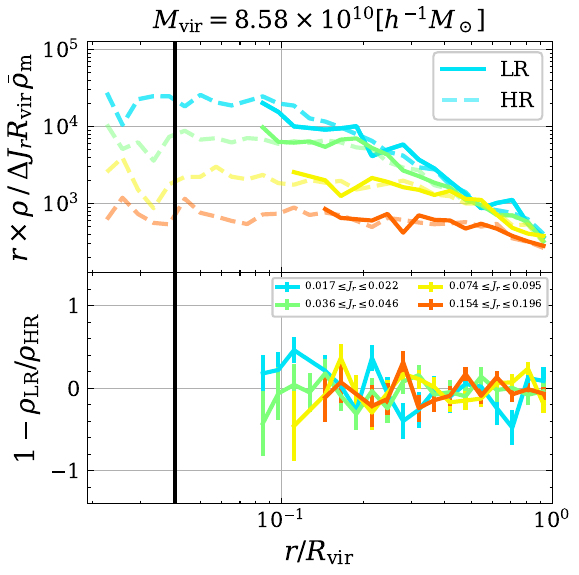}
        \end{minipage} &
        \begin{minipage}{0.66\columnwidth}
        \centering            \includegraphics[scale=0.56]{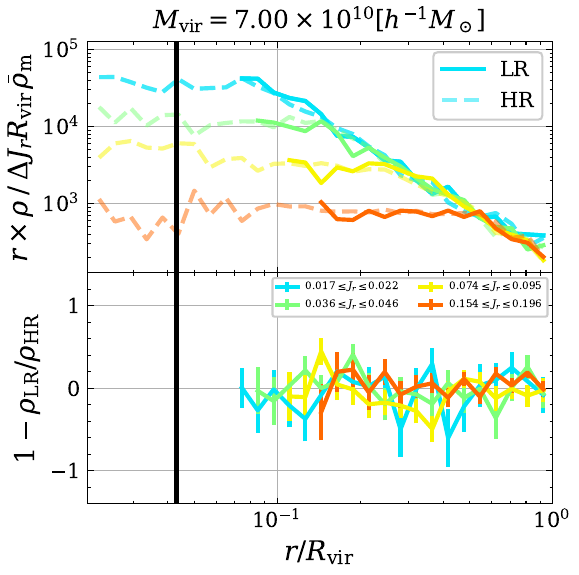}
        \end{minipage} &
        \begin{minipage}{0.66\columnwidth}
        \centering            \includegraphics[scale=0.56]{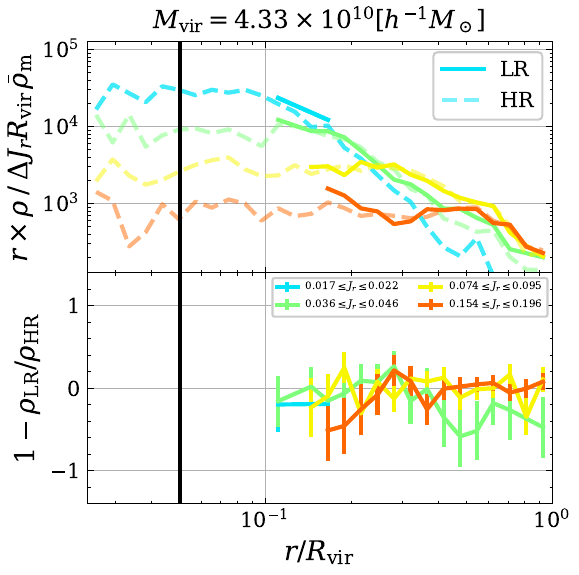}
        \end{minipage}
    \end{tabular}
    \caption{Comparison between the density profiles of particles classified by radial action $J_r$ of representative halos in WDM05 (denoted as LR, solid lines) and WDM05HR (denoted as HR, dashed lines) simulations. 
    In every panel, we show density profiles in the top panel and the fractional difference between them in the bottom panel.
    The error bars in the bottom panels are computed as Poisson shot noise.
    For profiles of the WDM05 simulation, we exclude bins containing less than $10$ particles.
    The vertical black lines indicate three times the softening length of the WDM05 simulation, and the leftmost radius shown is that of WDM05HR.
    We used the position and the velocity of halos taken from the ROCKSTAR halo catalog to calculate radial action in each simulation.
    Alt Text: Comparison of the density profiles in the $0.5$ keV WDM model between the high and low-resolution simulations, providing that our results in section~\ref{section:results} are less affected by numerical fragmentation.
    }
    \label{fig:Jrdens_massres}
\end{figure*}
%%%%%%%%%%%%%%%%%%%%%%%%%%%%%%%%%%%%%%%%%%%%%%%%%%%%%%%%%%%%%%%%%%

In this appendix, we demonstrate that the density profiles presented in section~\ref{section:results} are unaffected by the mass resolution of the $N$-body simulations. To illustrate this, we compare the profiles with those obtained from a higher-resolution simulation. Specifically, we conduct an additional simulation of the $0.5$ keV WDM model using the same initial conditions as the WDM05 simulation but with an increased particle number of $1000^3$ and twice better force resolution than the WDM05 simulation.
This simulation is referred to as WDM05HR in what follows. Since the only difference between these simulations is mass resolution, we can directly compare the profiles of the same halos across the two simulations.

In figure~\ref{fig:Jrdens_massres}, we compare $\rho(r;J_r)$ of representative halos between WDM05 and WDM05HR simulations to confirm the convergence of our results.
As is clearly shown from the bottom row of each panel, the profiles in the WDM05 simulations are consistent with those in the WDM05HR simulation within error bars for most of the profiles shown.
Based on these results, we conclude that our results in section~\ref{section:results} are less affected by numerical fragmentation induced by the discreteness of $N$-body particles.
Finally, it should be noted that we can observe $\rho \propto r^{-1}$ feature in the WDM05HR profiles below the resolution limit of WDM05 simulation (vertical black solid lines).
It would be interesting to investigate whether that feature extends to $r \to 0$ limit utilizing higher resolution simulation, though this is beyond the scope of this paper.

%%%
% See the manual for the detail.
%%%

\bibliographystyle{apj}
\bibliography{202207library}

\end{document}